%% file: ms.tex
\documentclass[9pt,conference]{IEEEtran}

\usepackage{waspaa25} %

\usepackage{bm} %
\usepackage{amsmath,amssymb,amsfonts}
\usepackage{circuitikz}
\usepackage{subcaption}
\usepackage[dvipsnames]{xcolor}
\usepackage{tikz}
\usetikzlibrary{decorations.pathreplacing,calligraphy, arrows.meta}
\usepackage{pgf}
\usepackage{pgfplots}
\usetikzlibrary{intersections}
\usetikzlibrary{positioning}
\usetikzlibrary{patterns}
\usepgfplotslibrary{fillbetween}
\usepackage{pifont}
\usepackage[nolist, nohyperlinks]{acronym}
\usepackage{nicefrac}
\usepackage{hyperref}
\usepackage{cleveref}
\usepackage{flushend}
\usepackage{cite}
\pgfplotsset{compat=1.18}
\usepackage{csquotes}
\usepackage{xspace}

\acrodefplural{nns}[NNs]{neural networks}
\acrodefplural{irm}[IRMs]{ideal ratio masks}
\acrodefplural{ibm}[IBMs]{ideal binary masks}
\acrodefplural{doa}[DoAs]{directions of arrival}
\acrodefplural{rirs}[RIRs]{room impulse responses}

\begin{acronym}
\acro{stft}[STFT]{short-time Fourier transform}
\acro{rir}[RIR]{room impulse response}
\acro{istft}[iSTFT]{inverse short-time Fourier transform}
\acro{doa}[DoA]{direction of arrival}
\acro{irm}[IRM]{ideal ratio mask}
\acro{mvdr}[MVDR]{minimum-variance distortionless response}
\acro{nn}[NN]{neural network}
\acro{dnn}[DNN]{deep neural network}
\acro{gcc_phat}[GCC-PHAT]{generalized cross-correlation phase transform}
\acro{ssl}[SSL]{sound source localization}
\acro{srp_phat}[SRP-PHAT]{steered response power with phase transform}
\acro{mse}[MSE]{mean squared error}
\acro{mae}[MAE]{mean angular error}
\acro{sdr}[SDR]{signal-to-distortion ratio}
\acro{sisdr}[SI-SDR]{scale-invariant signal-to-distortion ratio}
\acro{pesq}[PESQ]{perceptual evaluation of speech quality}
\acro{wer}[WER]{word error rate}
\acro{pit}[PIT]{permutation-invariant training}
\acro{lbt}[LBT]{location-based training}
\acro{asr}[ASR]{automatic speech recognition}
\acro{bce}[BCE]{binary cross-entropy}
\acro{jnf}[JNF]{joint non-linear filter}
\acro{nn}[NN]{neural network}
\acro{dnn}[DNN]{deep neural network}
\acro{mccruse}[MC-CRUSE]{multi-channel convolutional recurrent U-net architecture for speech enhancement}
\acro{estoi}[ESTOI]{extended short-time objective intelligibility}
\acro{snr}[SNR]{signal-to-noise ratio}
\acro{sir}[SIR]{signal-to-interference ratio}
\acro{iid}[i.i.d.]{independent and identically distributed}
\acro{wrt}[w.r.t.]{with respect to}
\acro{ssf}[SSF]{spatially selective filter}
\acro{cv}[CV]{constant velocity}
\acro{rw}[RW]{random walk}
\acro{ar}[AR]{autoregressive}
\acro{tse}[TSE]{target speaker extraction}
\acro{tst}[TST]{target speaker tracking}
\acro{tsl}[TSL]{target speaker localization}
\acro{dnsmos}[DNSMOS]{deep noise suppression mean opinion score}
\acro{map}[MAP]{maximum a posteriori}
\acro{kf}[KF]{Kalman filter}
\acro{pf}[PF]{particle filter}
\acro{das}[DaS]{delay-and-sum}
\acro{ae}[AE]{angular error}
\acro{acc}[ACC]{accuracy}
\acro{miso}[MISO]{multiple-input and single-output}
\acro{mimo}[MIMO]{multiple-input and multiple-output}
\end{acronym}

\newcommand\tickWidth{0.5pt}
\newcommand\rectEdge{1mm}

\newcommand\idxParticles{n}
\newcommand\numParticles{N}
\newcommand\perturb{\eta}

\usetikzlibrary{shapes.misc, arrows.meta, shapes.geometric}
\tikzset{cross/.style={cross out, draw=black, minimum size=2*(#1-\pgflinewidth), inner sep=0pt, outer sep=0pt},}

\definecolor{tab_blue}{RGB}{31, 119, 180}
\definecolor{tab_green}{RGB}{44, 160, 44}
\definecolor{tab_orange}{RGB}{255, 127, 14}
\definecolor{tab_purple}{RGB}{148, 103, 189}
\definecolor{tab_brown}{RGB}{165, 42, 38}
\definecolor{tab_pink}{RGB}{227, 119, 194}
\definecolor{tab_cyan}{RGB}{23, 190, 207}
\definecolor{orig_green}{RGB}{44, 160, 44}
\definecolor{orig_purple}{RGB}{148, 103, 189}
\definecolor{custom_green}{RGB}{149.5, 207.5, 149.5}
\definecolor{custom_purple}{RGB}{201.5, 179, 222}
\definecolor{dark_gray}{RGB}{128, 128, 128}

\DeclareMathAlphabet\mathbfcal{OMS}{cmsy}{b}{n}

\if@doubleblind
        \title{Self-Steering Deep Non-Linear Spatially Selective Filters for Efficient Extraction of Moving Speakers under Weak Guidance
    \thanks{This is a placeholder for sponsor acknowledgments. \\
    \phantom{German Research Foundation) under grant 508337379 and Erlangen National} \\
    \phantom{High Performance Computing Center (NHR@FAU) of the Friedrich-Alexander-} \\
    \phantom{Universität Erlangen-Nürnberg (FAU) under the NHR project ??????. NHR} \\
    \phantom{funding is provided by federal and Bavarian state authorities. NHR@FAU hardware}\\
    \phantom{is partially funded by the DFG under grant 440719683.}
    }}
\else
    \title{Self-Steering Deep Non-Linear Spatially Selective Filters for Efficient Extraction of Moving Speakers under Weak Guidance
    \thanks{This work was supported by the Deutsche Forschungsgemeinschaft (DFG, German Research Foundation) under grant 508337379 and Erlangen National High Performance Computing Center (NHR@FAU) of the Friedrich-Alexander-Universität Erlangen-Nürnberg (FAU) under the NHR project f104ac. NHR funding is provided by federal and Bavarian state authorities. NHR@FAU hardware is partially funded by the DFG under grant 440719683.}}
\fi

\name{Jakob Kienegger,
      Alina Mannanova,
      Huajian Fang,
      Timo Gerkmann}
\address{University of Hamburg, Germany}

\begin{document}

\maketitle

\begin{abstract}
Recent works on deep non-linear spatially selective filters demonstrate exceptional enhancement performance with computationally lightweight architectures for stationary speakers of known directions.
However, to maintain this performance in dynamic scenarios, resource-intensive data-driven tracking algorithms become necessary to provide precise spatial guidance conditioned on the initial direction of a target speaker.
As this additional computational overhead hinders application in resource-constrained scenarios such as real-time speech enhancement, we present a novel strategy utilizing a low-complexity tracking algorithm in the form of a \acl{pf} instead.
Assuming a causal, sequential processing style, we introduce temporal feedback to leverage the enhanced speech signal of the \acl{ssf} to compensate for the limited modeling capabilities of the \acl{pf}.
Evaluation on a synthetic dataset illustrates how the \acl{ar} interplay between both algorithms drastically improves tracking accuracy and leads to strong enhancement performance.
A listening test with real-world recordings complements these findings by indicating a clear trend towards our proposed self-steering pipeline as preferred choice over comparable methods.
\end{abstract}

\section{Introduction}
The field of speech enhancement focuses on improving the quality and intelligibility of a recorded speech signal by suppressing noise and reverberation artifacts.
Additional interfering speakers present a particularly challenging type of noise due to their similar and non-stationary statistical characteristics.
In such a multi-speaker environment, the task of \ac{tse} lies in identifying and enhancing the speech signal of a specific target speaker based on supplementary information, also known as cues.
Although a range of cue modalities have been suggested for this purpose \cite{zmolikova23tse_overview}, spatial information becomes especially popular when recordings from a microphone array are available.
Commonly under the assumption that contributing speakers are distinguishable by their azimuth orientation to the array, a \ac{ssf} can be employed and steered in the target's direction to extract the desired speech signal. 
Compared to direct separation approaches disentangling all speech sources at once, \acp{ssf} have been shown to be superior in the enhancement task \cite{tesch24ssf_journal}.
Especially in scenarios with stationary speakers, deep non-linear \acp{ssf} demonstrate exceptional performance while retaining lightweight \ac{nn} architectures \cite{tesch24ssf_journal, briegleb23icospa, bohlender24sep_journal, pandey12directional_speech_extraction, gu24rezero}.

A handful of constrained recording setups, such as a seated conference meeting with a centered microphone array \cite{chen20libricss}, may legitimate the assumption of stationary and directionally distinct speaker locations. However, more general settings, like the dinner party scenario considered in \cite{barker18chime5}, leave them unquestionably invalid.
Due to utilizing temporal context in the extraction process \cite{tesch24ssf_journal, briegleb23icospa, bohlender24sep_journal, pandey12directional_speech_extraction, gu24rezero}, the change from a stationary to a dynamic acoustic scenario results in a significant mismatch for data-driven \acp{ssf}.
To minimize performance degradation, speaker movement has to be included during training \cite{kienegger25wg_ssf}.
The arising need for time-dependent ground truth directional cues, which we refer to as \textit{strong} guidance, bears an additional challenge, since they are usually not available.
To increase applicability, \textit{weakly} guided \ac{tse} relies only on knowledge of the speakers starting position by including an upstream tracking algorithm to automate the steering of the \ac{ssf}.
Although a joint training strategy can increase robustness to cue inaccuracies, previous work has shown that a resource-intensive tracking algorithm remains necessary to provide sufficiently precise directional guidance and retain strong enhancement performance \cite{kienegger25wg_ssf}.

While the majority of speech enhancement systems conduct offline processing, an increasing demand in telecommunication, assistive technologies and consumer electronics boosts the research towards real-time capable solutions \cite{tan18cruse, defossez20demucs, braun21cruse}.
Since they are often implemented as low-complexity and causal modifications of their offline counterparts, such approaches tend to suffer from a fundamental disadvantage due to the reduced computational capacity \cite{luo19conv_tasnet, chao24SEmamba}. 
However, the sequential processing style also opens new possibilities such as improving the enhancement performance by utilizing the extracted speech signal from prior steps.
Recent works demonstrate promising results by introducing a temporal feedback to obtain \ac{ar} \ac{nn} architectures \cite{andreev23iterative_autoregression, pan24paris_autoregressive_separation}. 
Nevertheless, the inherent non-parallelizability during training limits their applicability in data-driven speech enhancement.

Instead of relying on computationally heavy methods utilizing \acp{nn} to guide the \ac{ssf}, we present a novel strategy to improve the accuracy of a low-complexity tracking algorithm in form of a \ac{pf}.
Assuming a sequential processing style common to real-time systems, we propose to utilize the enhanced speech of the \ac{ssf} via a temporal feedback to compensate for the limited modeling capabilities of the \ac{pf}. 
To keep the training effort light, we optimize the algorithms independently and only use a limited number of fine-tuning steps to train the resulting \ac{ar} architecture in an end-to-end fashion.
By evaluating our proposed \textit{self-steering} pipeline with a synthetic dataset, we can demonstrate a drastic improvement in tracking accuracy leading to strong enhancement performance. 
Complemented by a listening test with real-world recordings, our method achieves superior speech quality and interference suppression towards comparable approaches. 

\section{Signal Model}\label{sec:problem_definition}
We model the multi-channel observation $\mathbf{Y}$ as mixture between anechoic target speech signal $\mathbf{S}$ and noise $\mathbf{V}$.
The latter contains interfering speech signals, additive environmental noise as well as the reverberant components of the target speech.
In the \ac{stft} domain, this acoustic setup is resembled by
\begin{equation}\label{eq:signal_model}
    \mathbf{Y}_{tk} = \mathbf{S}_{tk} + \mathbf{V}_{tk} \, ,
\end{equation}
with $t$ and $k$ denoting frame and frequency bins respectively and vectorization conducted over the channel dimension.
In this work we set the goal of reconstructing the anechoic target speech signal $\mathbf{S}_{tk}$ at a predefined reference microphone, which we indicate as $S_{tk}$.
Thus, the task of \ac{tse} lies in recovering the speech signal
$S_{tk}$ from the noisy observations $\mathbf{Y}_{tk}$ conditioned on a speaker specific characteristic.

\section{Spatially Guided Speaker Extraction}
\subsection{Strongly guided target speaker extraction}\label{sec:strong_tse}
Spatially selective filters (\acp{ssf}) exploit positional cues to distinguish the target's speech signal from interfering sources. 
In this work, we abide by the common convention of restraining the cue modality to the relative azimuth orientation $\theta_t$, referred to as \ac{doa}, of the target speaker to the microphone array \cite{tesch24ssf_journal, briegleb23icospa, bohlender24sep_journal}.
If ground truth \ac{doa} information is available, a \ac{ssf} can be directly employed to solve the extraction task, which we refer to as strongly guided \ac{tse}.
A common approach to recover the target speech signal is to compute a time-frequency mask $\mathcal{M}_{tk}$ and extract $S_{tk}$ by multiplication with the corresponding reference channel $Y^0_{tk}$ of observation $\mathbf{Y}_{tk}$ \cite{tesch24ssf_journal, bohlender24sep_journal},
\begin{equation}\label{eq:mask_extraction}
    \widehat{S}_{tk} = \mathcal{M}_{tk} Y^0_{tk} \, .
\end{equation}
In terms of sequential processing, the \ac{ssf} may thus be seen as a deterministic function $\mathcal{F}$ mapping current broadband input $\mathbf{Y}_{t}$ and \ac{doa} $\theta_t$ to mask values $\mathcal{M}_{tk}$.
Due to utilizing temporal dependencies \cite{tesch24ssf_journal, briegleb23icospa, bohlender24sep_journal, pandey12directional_speech_extraction, gu24rezero}, deep \acp{ssf} are parametrized by a hidden state $\mathbf{z}_{t-1}$ capturing previous context in addition to the parameters $\phi$ of the \ac{nn} architecture
\begin{equation}\label{eq:miso_mask_computation}
    \mathcal{M}_{tk} = \mathcal{F}_k\!\left( \mathbf{Y}_{t}, \theta_{t} ; \mathbf{z}_{t-1}, \phi \right) \, .
\end{equation}
Regarding the channel dimension, the \ac{ssf} may thus be seen as a \ac{miso} system by computing a single-channel mask $\mathcal{M}_{tk}$ based on multi-channel observations $\mathbf{Y}_{t}$. 

\subsection{Weakly guided target speaker extraction}
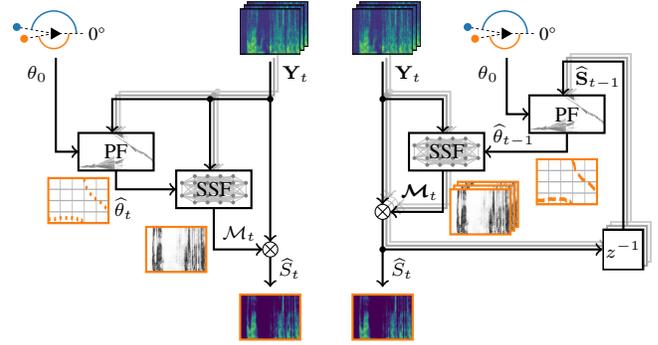
\begin{figure}[t!]
\centering
\begin{subfigure}{0.49\linewidth}
    \input{images/weak_pipeline/weak_pipeline.tikz}
     \caption{Concatenation of tracking and enhancement algorithms.}
     \label{fig:weak_pipeline}
 \end{subfigure}
 \hfill
 \begin{subfigure}{0.49\linewidth}
     \input{images/self_pipeline/self_pipeline.tikz}
     \caption{Proposed \acl{ar} (\acs{ar}) integration of both algorithms.}
     \label{fig:self_pipeline}
 \end{subfigure}
    \caption{Weakly guided \acl{tse} (\acs{tse}) frameworks which interconnect a \acl{pf} (\acs{pf}) as tracking algorithm with a deep non-linear \acl{ssf} (\acs{ssf}) for speech enhancement. }
    \label{fig:pipeline_overview}
\end{figure}

The dependency of strongly guided \ac{tse} on the continuous provision of ground truth directional information poses a great limitation in most practical scenarios.
As an alternative, weakly guided \ac{tse} \cite{kienegger25wg_ssf} restrains the cue only to the target's initial orientation $\theta_0$ \acs{wrt} the microphone array.
In order to continue using a \ac{ssf} for the enhancement task, an upstream tracking algorithm predicting the temporal evolution of the \ac{doa} $\theta_t$ becomes necessary.
Due to its well-balanced tradeoff between computational complexity and robustness, we choose a \ac{pf} for this task, which has already proven highly effective in the context of acoustic speaker tracking \cite{vermaak01pf_nonlinear_speaker_tracking, ward03basic_particle_filter, dong20pf_doa_coprime}.
Modeling the tracking problem in a state-space formulation, the \ac{pf} recovers the \ac{doa} $\theta_t$ encoded in state $\mathbf{x}_t$ from all available observations $\mathbf{Y}_{1:t}$ and based on the initial condition $\mathbf{x}_0$.  
The state $\mathbf{x}_t$ contains the \ac{doa} $\theta_t$ either explicitly or implicitly in a different coordinate representation and is often complemented by velocity or acceleration for an improved modeling of motion dynamics.
A \ac{pf} progressively updates a discrete approximation of the posterior $p(\mathbf{x}_t|\mathbf{Y}_{1:t}, \mathbf{x}_0)$ using sequential importance sampling \cite{hagiwara21state_space_model}
\begin{equation}\label{eq:sis}
    p(\mathbf{x}_t|\mathbf{Y}_{1:t}, \mathbf{x}_0) \approx\sum_{\idxParticles=1}^\numParticles w_t^{\scriptscriptstyle(\idxParticles)} \delta\!\left(\mathbf{x}_t\!-\!\mathbf{x}_t^{\scriptscriptstyle(\idxParticles)}\right) \, ,
\end{equation} 
with $\delta(\cdot)$ denoting the Dirac delta function.
The $\numParticles$ state samples $\mathbf{x}_t^{\scriptscriptstyle(\idxParticles)}$, referred to as particles, are drawn according to a proposal distribution $q\!\left(\mathbf{x}_t|\mathbf{Y}_{1:t}, \mathbf{x}_{0:t-1}^{\scriptscriptstyle(\idxParticles)}\right)$ and the importance weights $w_t^{\scriptscriptstyle(\idxParticles)}$ can be recursively computed by imposing mild conditions on the underlying statistical model \cite{arulampalam02pf_tutorial, chen04pf_sensor_fusion}. 
Utilizing the approximated probability density function in \eqref{eq:sis} enables a tractable estimation of the posterior mean, which is used as the state estimate $\widehat{\mathbf{x}}_t$.
The specific \ac{pf} formulation referred to as bootstrap filter \cite{gordon93pf_bootstrap_filter} allows for a particularly efficient implementation by only taking the state-transition model for the proposal distribution into account. 
Although the resulting state estimates lack information from current observations, the algorithmic simplicity makes it a popular choice for acoustic tracking tasks \cite{ward03basic_particle_filter, dong20pf_doa_coprime}.
By leveraging the state estimates of a \ac{pf} to guide a downstream \ac{ssf} based on observations $\mathbf{Y}_{t}$ and starting direction $\theta_0$, the \ac{tse} pipeline emerges as concatenation of both algorithms, see \cref{fig:weak_pipeline}.

\begin{figure*}[t!]
\input{images/trajectories/trajectories.tikz}
\vspace*{-15pt}
\caption{
Synthetic two-speaker ({\protect\tikz[baseline=-1ex]  \protect\draw[tab_blue, line width=1pt] (0,0mm) -- (1mm,0mm);}/{\protect\tikz[baseline=-0.33ex] \protect\draw[tab_orange, line width=1pt] (0,0mm) -- (1mm,0mm);}) dataset with continuous movement. The motion model reflects natural human walking speed\cite{keller96walking_jogging_running_speed, murtagh21outdoor_walking_speed}. 
}
\label{fig:trajectories}
\end{figure*}
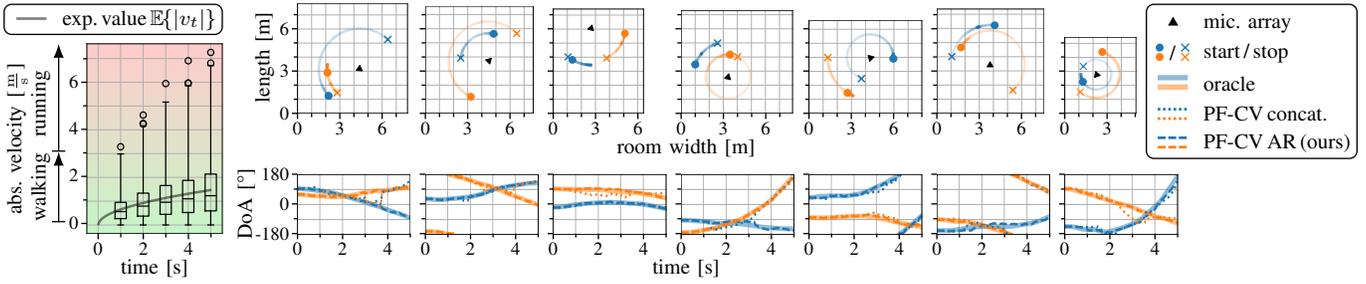

\section{Proposed method}
The sequential application of tracking and enhancement algorithms may seem like the natural approach to automate the steering of a \ac{ssf}.
While this strategy has proven successful in our previous work on \ac{tse} \cite{kienegger25wg_ssf} as well as for the related task of speech separation in \cite{bohlender24sep_journal},
it is overly dependent on the accuracy of the upstream \ac{doa} estimates.
As a result, computationally heavy, data-driven tracking algorithms have to be employed to provide sufficiently precise directional information and maintain a strong enhancement performance \cite{bohlender24sep_journal, kienegger25wg_ssf}.
This additional computational overhead naturally limits the applicability of such methods in resource-constrained scenarios. 
Instead of an independent concatenation, we propose a more dynamic interaction between tracking and enhancement algorithms. 
Since the nature of sequential processing grants access to the output of the \ac{ssf}, we introduce a temporal feedback to improve the accuracy of a computationally lightweight \ac{pf} as opposed to employing a resource-intensive tracking algorithm.
To retain spatial information at the output of the \ac{ssf}, we slightly deviate from the original formulation in \eqref{eq:mask_extraction}.
Rather than just for the reference channel, we propose to compute a multi-channel mask $\mathbfcal{M}_{tk}$ and extract the direct path propagated target speech signal $\mathbf{S}_{tk}$ by element-wise multiplication ($\odot$) with the observation $\mathbf{Y}_{tk}$
\begin{equation}\label{eq:mimo_mask_extraction}
    \widehat{\mathbf{S}}_{tk} = \mathbfcal{M}_{tk} \odot \mathbf{Y}_{tk} \, .
\end{equation}
Unlike the original \ac{miso} formulation in \eqref{eq:mask_extraction}, the estimates $\widehat{\mathbf{S}}_{tk}$ of such a \ac{mimo} strategy have already been proven to result in an improved localization performance for stationary scenarios \cite{li23gcc-speaker, chen24locselect, battula25MIMO_localization}. 
Rather than approximating the posterior based on the broadband observations $\mathbf{Y}_{t}$ containing the directional information of all speakers, the \ac{pf} algorithm can utilize the spatial characteristics of the target's direct path propagation
$p\big(\mathbf{x}_t|\mathbf{S}_{1:t}, \mathbf{x}_0\big)$.
However, due to the temporal feedback in the \ac{ar} structure, only the \ac{doa} estimates from the previous frames $\widehat{\theta}_{t-1}$ are accessible to the \ac{ssf}. 
This necessitates a directional single-frame lookahead prediction during mask computation, leading to the adjusted formulation
\begin{equation}\label{eq:mimo_mask_computation}
    \mathbfcal{M}_{tk} = \mathbfcal{F}_k\!\left( \mathbf{Y}_{t}, \theta_{t-1} ; \mathbf{z}_{t-1}, \phi \right) \, .
\end{equation}
Paired with the need to reconstruct the anechoic propagation between the microphones of the array, the difficulty in the enhancement task increases for the \ac{ssf} compared to its original formulation in \eqref{eq:miso_mask_computation}.
\Cref{fig:self_pipeline} displays our proposed weakly guided \ac{tse} pipeline as an \ac{ar} interconnection between \ac{pf} and our \ac{mimo} adaption of the \ac{ssf}.

\section{Experiments}
\subsection{Dataset}\label{sec:dataset}
\hspace{\the\parindent}\textit{\textbf{Synthetic dataset}}
We employ a synthetic dataset consisting of noisy and reverberant two-speaker mixtures to enable \ac{nn} training, model optimization and evaluation in a controlled acoustic scenario.  
In particular, we use utterances from the LibriSpeech corpus \cite{panayotov15librispeech} and pair them according to the recipe in Libri2Mix \cite{cosentino20librimix}.
For spatialization, we utilize simulated \acp{rir} in shoe-box shaped rooms based on a GPU accelerated implementation \cite{diaz18gpu_rir} of the image method \cite{allen79image_method}.
We closely follow the works from Tesch et al. \cite{tesch24ssf_journal, tesch22spatial_spectral_temporal_nonlinear_filter, tesch23deep_nonliner_filter_multichannel, tesch23ssf} by using the same three microphone circular array of $10\,\mathrm{cm}$ diameter and adhering to reverberation times between 0.2\,s and 0.5\,s.
While retaining the same randomized microphone placement and speaker height, we increase the distance between array and speakers to a range of 1\,m to 3\,m whilst maintaining a minimum distance of 1\,m to the surrounding walls and 15\,cm among the speakers themselves.
Building on our previous work \cite{kienegger25wg_ssf}, we model the speaker trajectories as circular motion with constant distance and height, allowing for a straightforward parametrization and intuitive understanding of the spatial movement patterns.
To include directional changes and random acceleration behavior, we choose the azimuthal motion model 
\begin{equation}\label{eq:cv_motion_model}
    \begin{bmatrix}
    \theta_{t+1} \\
    \dot{\theta}_{t+1}
    \end{bmatrix} = \begin{bmatrix}
        1 & \hspace{-2mm}\Delta T \\
        0 & 1
    \end{bmatrix} \begin{bmatrix}
    \theta_t \\
    \dot{\theta}_t
    \end{bmatrix} + \begin{bmatrix}
        \Delta T^2/ 2 \\
        \Delta T
    \end{bmatrix} \perturb_t \, ,
\end{equation}
referred to as \ac{cv} or white acceleration model \cite{rong03survey_target_tracking}.
The state vector $\mathbf{x}_t$ in this linear state-space formulation consists of $\begin{bmatrix}
    \theta_t, \dot{\theta}_t
\end{bmatrix}^\top$, thus, azimuth orientation and instantaneous azimuth velocity, respectively. The perturbation $\perturb_t$ is a white, zero mean Gaussian process of variance $\sigma^2$.
Due to the wide range of distances $r$ between speakers and microphone array, 
we use the inter-frame velocity $v_t$ defined as $\frac{r}{\Delta T}(\theta_t - \theta_{t-1})$ instead of the angular displacement in \cite{kienegger25wg_ssf} as a means to obtain a physically meaningful parametrization for the motion model.
Since the difference $\theta_t - \theta_{t-1}$ is not affected by wrapping effects, sequentially evaluating the linear state-space model in \eqref{eq:cv_motion_model} retains its Gaussianity.
This makes the expected absolute velocity $|v_t|$ at distance $r$ tractable and yields the closed-form expression
\begin{equation}\label{eq:expected_abs_vel}
    \mathbb{E}\hspace*{-1pt}\big\{ |v_t| \big| r \big\} = \Delta T \, r \sqrt{\frac{4t-3}{2\pi}} \sigma \, .
\end{equation} 
We will use this correspondence to determine the standard deviation $\sigma$ of the motion perturbation $\perturb_t$ in \eqref{eq:cv_motion_model} to reach an expected absolute velocity $|v_t|$ of 1.5\,{\small $\frac{\mathrm{m}}{\mathrm{s}}$} at trajectory end, which corresponds to medium walking speed \cite{keller96walking_jogging_running_speed, murtagh21outdoor_walking_speed}.
Due to the steady increase of the expected value in \eqref{eq:expected_abs_vel}, we will only use trajectories of 5\,s length and truncate the utterances accordingly.
\Cref{fig:trajectories} illustrates the resulting distribution of $|v_t|$ as well as a selection of simulation scenarios from the test set.
The step-size $\Delta T$ for the temporal discretization of the trajectories is aligned with the \ac{stft} parametrization, for which we use a square-root Hann window \cite{shimauchi14hann_window} of length $32\,\mathrm{ms}$ and $16\,\mathrm{ms}$ hop-size. 
After spatialization, we add spherical isotropic white noise according to \cite{habets07isotropic_noise_generation} with a \ac{snr} between 20\,dB and 30\,dB.

\textit{\textbf{Recorded dataset}}
To test generalizablity and robustness in real-world scenarios, we also include recordings with human speakers for evaluation.
We use a microphone array matching the training configuration in a room of $9.5\,\mathrm{m}\times 5.1\,\mathrm{m}\times 2.4\,\mathrm{m}$ extent and a reverberation time of approximately 0.35\,s.
Two male speakers are instructed to simultaneously read out segments from the Rainbow Passage \cite{fairbanks60rainbow_passage} while moving in a circular trajectory.
The speakers start from opposite ends at roughly 1\,m and 2\,m distance from the array, traverse to the other end of the room, and back during the duration of their prompt.
We conduct 3 recordings in total, ranging from 10\,s to 20\,s.
Thus, each scenario contains two speaker crossings with velocities depending on distance to the array and length of the prompt. 
To visualize the setup, we provide a video of two of the recordings as supplementary material: \texttt{https://youtu.be/aSKOSh5JZ3o} 

\subsection{Algorithm implementation and optimization details}\label{sec:nn_training}

\hspace{\the\parindent}\textit{\textbf{Spatially selective filter}} 
For sequential processing, recurrent \acp{nn} are particularly convenient as they store the temporal context in a compact hidden state.
Due to its purely recurrent nature combined with a conceptually simple and lightweight architecture, we choose FT-\acs{jnf} \cite{tesch24ssf_journal} as \ac{ssf}, which has seen great popularity in a variety of applications \cite{wechsler24directional_filtering_directivity_control, briegleb24constrained_vs_unconstained_filtering, lentz24ftjnf_head_rotation}.
Following our previous work \cite{kienegger25wg_ssf}, we slightly deviate from the implementation in \cite{tesch24ssf_journal} and modify the last B-LSTM layer to obtain a causal \ac{ssf} while retaining roughly the intended model size.
The modifications to compute a multi-channel mask according to our \ac{mimo} formulation in \eqref{eq:mimo_mask_computation} only account for a negligible increase of less than 0.1\,\% in model parameters and 0.05\,\% in GMACs.
We train FT-\acs{jnf} in both \ac{miso} and \ac{mimo} configurations for 300 epochs while adhering to the original scheduler with an initial learning rate of $10^{-3}$ and a joint time and frequency domain loss function \cite{tesch24ssf_journal}.
The latter is averaged over all channels in the \ac{mimo} formulation.
Based on our previous findings of improved robustness to tracking inaccuracies \cite{kienegger25wg_ssf}, we fine-tune the \acp{ssf} for weakly guided \ac{tse} in both concatenative and self-steering pipelines for 25 additional epochs with a reduced learning rate of $10^{-4}$.
Instead of using iterative approaches to avoid the non-parallelizable training of our \ac{ar} architecture \cite{andreev23iterative_autoregression, pan24paris_autoregressive_separation}, we found that the small amount of fine-tuning steps yielded the best performance while remaining computationally feasible.
Furthermore, we observed consistently improved convergence when including \ac{mae} as an auxiliary loss function \cite{li23gcc-speaker}.

\textit{\textbf{Particle filter}}
We choose the generic bootstrap filter algorithm in \cite{lehmann06pf_resampling_Neff} as low-complexity tracking framework, which introduces an additional conditional resampling step to counter weight degeneracy problems \cite{doucet01pf_practice}.
Although providing a zero gradient for frames affected by this resampling step, we found it yielding better results during fine-tuning of the \acp{ssf} than a differentiable resampling algorithm \cite{karkus18pf_differentiable_resampling}.
We use the output power of a \acl{das} beamformer to construct the likelihood \cite{ward03basic_particle_filter} and employ two different state-transition densities to investigate the effect of mismatch in motion dynamics. 
Thus, we include a simpler, angular \ac{rw} model \cite{traa13wrapped_kalman_filter} in addition to the \ac{cv} formulation in \eqref{eq:cv_motion_model} used in our synthetic dataset.
We fix the number of particles $N$ to 50 and optimize the remaining model parameters with an exhaustive search on the validation subset.
 
\section{Results}
\begin{table}[t!]
 \caption{
Tracking (\acs{ae}) and enhancement (\acs{sisdr}, \acs{pesq}) performance of different strongly and weakly guided \ac{tse} pipelines using FT-\acs{jnf} \cite{tesch24ssf_journal} as \ac{ssf}. Results are presented in 25\%$|$50\%$|$75\% quartiles. 
}
\label{tab:results}
\resizebox{\columnwidth}{!}{
\setlength{\tabcolsep}{2pt} %
\renewcommand{\arraystretch}{0.9}
\footnotesize
\begin{tabular}{clllcccc}
 \toprule[1.5pt]
 & \multicolumn{3}{c}{\textbf{Method}} & \multicolumn{3}{c}{\textbf{Performance}\,[25\%$|$50\%$|$75\%]}\\[1pt] \cmidrule(lr){2-4} \cmidrule(lr){5-7} 
\textbf{ID} & Enhancement & Tracking & Pipeline & \acs{ae}\,[°]\,$\downarrow$ & \acs{sisdr}\,[dB]\,$\uparrow$ & \acs{pesq}\,$\uparrow$ \\ \midrule
(0) & unprocessed & \multicolumn{1}{c}{$-$} & \multicolumn{1}{c}{$-$} & $-$ & -10.1$|$-7.62$|$-4.99 & 1.05$|$1.07$|$1.10 \\ [-1pt] \cmidrule(lr){1-7}
(1) & \acs{miso} \cite{tesch24ssf_journal} & oracle & \multicolumn{1}{c}{$-$} & $-$ & \textbf{2.34}$|$\textbf{4.44}$|$\textbf{6.64} & 1.52$|$1.75$|$2.05 \\
(2) & \acs{mimo}\,(ours) & oracle & \multicolumn{1}{c}{$-$} & $-$ & 2.07$|$4.12$|$6.35 & \textbf{1.54}$|$\textbf{1.77}$|$\textbf{2.06} \\ [-1pt] \cmidrule(lr){1-7}
(3) & \acs{miso} \cite{tesch24ssf_journal} & \acs{pf}-\acs{rw} & concat. & 7.05$|$13.9$|$40.0 & 0.28$|$2.72$|$5.23 & 1.28$|$1.46$|$1.71 \\
(4) & \acs{mimo}\,(ours) & \acs{pf}-\acs{rw} & \acs{ar}\,(ours) & \textbf{2.31}$|$\textbf{3.14}$|$\textbf{4.74} & \textbf{1.33}$|$\textbf{3.54}$|$\textbf{5.79} & \textbf{1.41}$|$\textbf{1.62}$|$\textbf{1.88} \\ [-1pt] \cmidrule(lr){1-7}
(5) & \acs{miso} \cite{tesch24ssf_journal} & \acs{pf}-\acs{cv} & concat. & 5.56$|$9.78$|$24.4 & 0.59$|$3.05$|$5.58 & 1.30$|$1.50$|$1.78 \\
(6) & \acs{mimo}\,(ours) & \acs{pf}-\acs{cv} & \acs{ar}\,(ours) & \textbf{2.08}$|$\textbf{2.88}$|$\textbf{4.29} & \textbf{1.39}$|$\textbf{3.52}$|$\textbf{5.79} & \textbf{1.41}$|$\textbf{1.61}$|$\textbf{1.87} \\
 \bottomrule[1.5pt]
\end{tabular}
}
\end{table}

\begin{figure}[b!]
    \centering
    \input{images/tracking_enhancement_correlation/tracking_enhancement_correlation.tikz}
    \vspace*{-15pt}
    \caption{Dependency of enhancement performance (\acs{sisdr}) on tracking accuracy (\acs{mae}). Solid lines display a log-linear regression model.}
    \label{fig:tracking_enhancement_correlation}
\end{figure}
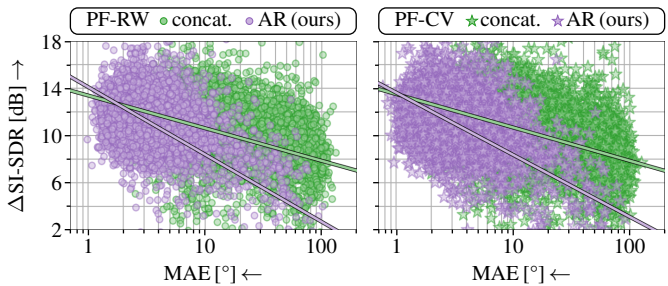
To perform a thorough assessment, we use our synthetic dataset for an \textit{objective} analysis and our recorded dataset for a \textit{subjective} evaluation conducted by human audio experts in a listening experiment. 
In both cases, the \ac{tse} methods are assessed regarding the perceptual speech quality of the enhanced signal and the \acl{sdr}.

\begin{figure}[t!]
\centering
\begin{subfigure}{0.49\linewidth}
    \hspace*{-5pt}\input{images/violin/violin.tikz}
     \vspace*{-13.1pt}
     \caption{Intrusive metrics representing speech quality (\acs{pesq}) and signal-to-distortion ratio (\acs{sisdr}).}
     \label{fig:violin_results}
     
 \end{subfigure}
 \hfill
 \begin{subfigure}{0.49\linewidth}
     \hspace*{-5pt}\input{images/preference_test/preference_test.tikz}
      \vspace*{-30pt}
     \caption{Preference listening test.}
     \label{fig:preference_test}
     
 \end{subfigure}
 
    \caption{Objective (a) and subjective (b) evaluation of speech enhancement performance on synthetic and recorded datasets respectively.}
    \label{fig:quality_distortions}
\end{figure}
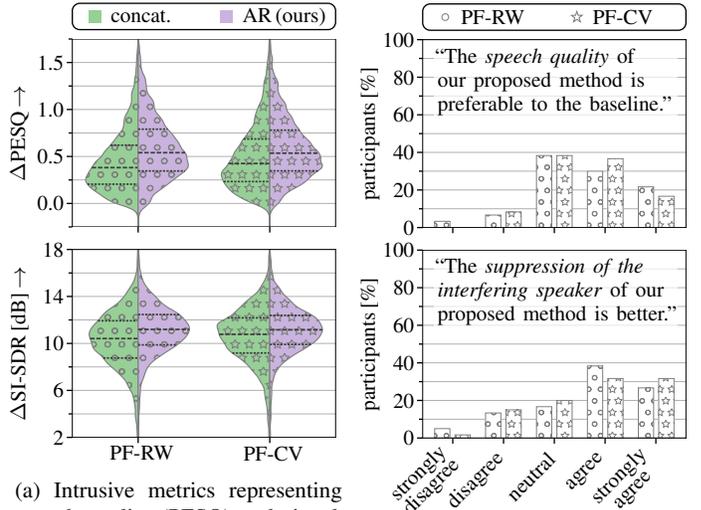

\textbf{\textit{Synthetic dataset}}
The availability of ground truth directional information allows the evaluation of the different \ac{ssf} formulations in a strongly guided fashion, eliminating the bias from upstream tracking inaccuracies.
\cref{tab:results} compares the original \ac{miso} (1) and our \ac{mimo} adaption (2) of FT-\acs{jnf} using the intrusive metrics \ac{pesq} \cite{rix01pesq} and \ac{sisdr} \cite{roux19sisdr_half_baked}.
While we are on par regarding speech quality, our proposed modifications result in a lower \acs{sisdr} score.
This can be attributed to the additional challenge from the delayed directional cue paired with the need of reconstructing the anechoic inter-microphone propagation.
Replacing the oracle \acp{doa} with estimated trajectories to evaluate the weakly guided methods, we complement the analysis by including frame-wise \ac{ae} to assess the accuracy of the \acp{pf}.
In their original formulation, both \ac{pf} versions display great inconsistency in their tracking performance (3,\,5). 
Especially the mismatched \ac{rw} motion model (3) yields a very poor accuracy with 25\,\% of the \ac{doa} estimations being more then 40° off.
By introducing our proposed \ac{ar} interconnection, we can drastically improve the performance of both \ac{pf} versions (4,\,6), as also demonstrated in the example trajectories in \cref{fig:trajectories}. 
\Cref{tab:results} and \cref{fig:violin_results} show how our method is able to translate the refined tracking into a superior speech enhancement performance compared with the concatenative approach for both perceptual quality and \acl{sdr} criteria (4,\,6).
To investigate the dependency of the two \ac{tse} methods on the precision of the directional cues, \cref{fig:tracking_enhancement_correlation} shows the relative \ac{sisdr} improvement in relation to utterance-wise \ac{mae}.
By indicating a stronger tie between tracking and enhancement metrics, it becomes apparent that our self-steering framework relies more heavily on \textit{spatial} information during the extraction process as opposed to the concatenative method. 
During fine-tuning, the latter gains robustness against the imprecise spatial conditioning and learns to leverage \textit{temporal-spectral} rather than spatial information \cite{kienegger25wg_ssf}.

\textbf{\textit{Recorded dataset}}
Motivated by the unavailability of clean speech signals in the recorded dataset, a listening test with human participants rating the enhancement performance is conducted.
We design the evaluation similar to the preference test in \cite{tesch24ssf_journal} by asking questions of the form \enquote{Example 1 is preferable over example 2} regarding speech quality and \acl{sdr}.
A total of 10 audio experts judge the enhancement performance regarding all 6 speech signals in the 3 recordings for both \ac{pf} versions, totaling to 12 test cases.
The results of the listening experiment in \cref{fig:preference_test} indicate a clear tendency towards our proposed \ac{ar} approach for both versions of the \ac{pf}. 
A closer investigation unveils that the most significant performance gain occurs for the speaker with greater distance to the array, underlining the benefit of utilizing spatial information in low \ac{snr} scenarios.

\section{Conclusion}\acresetall
In this work we proposed a novel weakly guided \ac{tse} method to improve resource-constrained speech enhancement in dynamic scenarios.
Under the assumption of a sequential processing style common to real-time systems, we suggested leveraging the enhanced speech of a \ac{ssf} via an \textit{\acl{ar}} approach to increase the accuracy of a low-complexity tracking algorithm in form of a \ac{pf}.
The resulting self-steering pipeline drastically improved the accuracy of the \ac{pf} and resulted in a strong enhancement performance.
A preference listening test with real-world recordings complemented these findings by outperforming comparable frameworks under the same conditions.

\bibliographystyle{IEEEtran}
\bibliography{refs, strings}

\end{document}

%% file: images/weak_pipeline/weak_pipeline.tikz
\begin{tikzpicture}

\definecolor{tab_blue}{RGB}{31,119,180}
\definecolor{tab_orange}{RGB}{255,127,14}
\definecolor{tab_green}{RGB}{44,160,44}

\newcommand\plotStartX{0mm}
\newcommand\plotStartY{0mm}

\newcommand\channelOffset{0.5mm}
\newcommand\pfOff{0mm}
\newcommand\specHeight{6mm}
\newcommand\specOff{1mm}
\newcommand\specWidth{8mm}
\newcommand\spkNum{1}
\ifthenelse{\spkNum=0}{\def\spkColor{tab_blue}}{\def\spkColor{tab_orange}}
\newcommand\exNum{3}
\newcommand\initDoaOff{-1.5mm}

\newcommand\arrowWidth{0.8pt}
\newcommand\arrowScale{1.5}
\newcommand\arrowR{2mm}
\newcommand\arrowLen{5mm} 
\newcommand\arrowCircleWidth{1.2pt}
\newcommand\crossR{1mm}
\pgfmathsetmacro{\crossOff}{\crossR * cos(45)}
\newcommand\zWidth{5mm}

\newcommand\largeFont{\footnotesize}
\newcommand\smallFont{\scriptsize}

\newcommand\boxHeight{5mm}
\newcommand\boxWidth{10mm}
\newcommand\boxCorner{4pt}
\newcommand\boxOff{3mm}

\pgfmathsetmacro{\recWidth}{
    2*\boxOff + 2*\boxWidth + 2*\arrowLen
}
\foreach \x in {1, 0, -1} {
    \pgfmathsetmacro{\absX}{abs(\x)}

    \ifthenelse{\x > 0}{\def\theColor{gray!40}}{}
    \ifthenelse{\x < 0}{\def\theColor{black}}{}
    \ifthenelse{\x = 0}{\def\theColor{gray!70}}{}
    
    \ifthenelse{\x < 0}{
    \draw[line width=\arrowWidth, color=\theColor, ->, >=to] (\plotStartX + \x * \channelOffset, \plotStartY+ \x * \channelOffset) -- (\plotStartX + \x * \channelOffset, \plotStartY - 4*\arrowLen - \boxHeight + \crossR+ \x * \channelOffset);
    \fill[\theColor] (\plotStartX + \x * \channelOffset, \plotStartY - \arrowLen+ \x * \channelOffset) circle (\arrowCircleWidth);
    \draw[line width=\arrowWidth, color=\theColor, ->, >=to] (\plotStartX + \x * \channelOffset, \plotStartY - \arrowLen+ \x * \channelOffset) -- (\plotStartX + \x * \channelOffset - \boxOff - 0.5 * \boxWidth, \plotStartY - \arrowLen+ \x * \channelOffset) -- (\plotStartX + \x * \channelOffset - \boxOff - 0.5 * \boxWidth, \plotStartY - 3*\arrowLen);
    }{
    \draw[line width=\arrowWidth, color=\theColor, ->, >=to] (\plotStartX + \x * \channelOffset, \plotStartY+ \x * \channelOffset) -- (\plotStartX + \x * \channelOffset, \plotStartY - \arrowLen+ \x * \channelOffset) -- (\plotStartX + \x * \channelOffset - \boxOff - 0.5 * \boxWidth, \plotStartY - \arrowLen+ \x * \channelOffset) -- (\plotStartX + \x * \channelOffset - \boxOff - 0.5 * \boxWidth, \plotStartY - 3*\arrowLen);
    }
    
 \fill[\theColor] (\plotStartX + \x * \channelOffset - \boxOff - 0.5 * \boxWidth, \plotStartY - \arrowLen+ \x * \channelOffset) circle (\arrowCircleWidth);
    \draw[line width=\arrowWidth, color=\theColor, ->, >=to] (\plotStartX + \x * \channelOffset - \boxOff - 0.5 * \boxWidth, \plotStartY - \arrowLen+ \x * \channelOffset) -- (\plotStartX + \x * \channelOffset - 2*\boxOff - 1.5 * \boxWidth, \plotStartY - \arrowLen+ \x * \channelOffset) -- (\plotStartX + \x * \channelOffset - 2*\boxOff - 1.5 * \boxWidth, \plotStartY - 2*\arrowLen);

    \ifthenelse{\x < 0}{
   \draw[line width=\arrowWidth, color=\theColor, ->, >=to] (\plotStartX  - \boxOff - 0.5 * \boxWidth, \plotStartY - 3*\arrowLen - \boxHeight) -- (\plotStartX - \boxOff - 0.5 * \boxWidth, \plotStartY -4*\arrowLen - \boxHeight + \x * \channelOffset) -- (\plotStartX + \x * \channelOffset - \crossR, \plotStartY - 4*\arrowLen - \boxHeight + \x * \channelOffset);
    }{
    } 

    \ifthenelse{\x < 0}{
        \draw[line width=\arrowWidth, color=\theColor, ->, >=to] (\plotStartX + \x * \channelOffset, \plotStartY - 4*\arrowLen - \boxHeight - \crossR + \x * \channelOffset) -- (\plotStartX + \x * \channelOffset, \plotStartY - 5*\arrowLen - \boxHeight+ \x * \channelOffset);
    }{
    }

    \ifthenelse{\x < 0}{
    \fill[color=white] (\plotStartX+ \x * \channelOffset, \plotStartY - 4*\arrowLen - \boxHeight+ \x * \channelOffset) circle (\crossR);
    \draw[color=\theColor] (\plotStartX+ \x * \channelOffset, \plotStartY - 4*\arrowLen - \boxHeight+ \x * \channelOffset) circle (\crossR);
    \draw[color=\theColor] (\plotStartX + \crossOff + \x * \channelOffset, \plotStartY - 4*\arrowLen - \boxHeight+ \crossOff+ \x * \channelOffset) -- (\plotStartX - \crossOff+ \x * \channelOffset, \plotStartY - 4*\arrowLen - \boxHeight- \crossOff+ \x * \channelOffset);
    \draw[color=\theColor] (\plotStartX - \crossOff+ \x * \channelOffset, \plotStartY - 4*\arrowLen - \boxHeight+\crossOff+ \x * \channelOffset) -- (\plotStartX  +\crossOff+ \x * \channelOffset, \plotStartY - 4*\arrowLen - \boxHeight -\crossOff+ \x * \channelOffset);
    }{
    }
}

\pgfmathsetlengthmacro{\w}{\boxWidth-\arrowWidth}
\pgfmathsetlengthmacro{\h}{\boxHeight-\arrowWidth}
\node[anchor=center] at (\plotStartX - 2*\boxOff - 1.5*\boxWidth, \plotStartY - 2*\arrowLen - 0.5*\boxHeight) {
    \pgfimage[width=\w, height=\h]{images/self_pipeline/particles_\exNum_\spkNum.png} 
};
\draw[line width=\arrowWidth] (\plotStartX - 2*\boxOff - 2*\boxWidth, \plotStartY - \boxHeight - 2*\arrowLen) rectangle (\plotStartX - 2*\boxOff - \boxWidth, \plotStartY - 2*\arrowLen);
\node[anchor=center] at (\plotStartX  - 2*\boxOff - 1.5*\boxWidth, \plotStartY - 2*\arrowLen - 0.5*\boxHeight) {
    \largeFont PF
};

\node[anchor=center] at (\plotStartX - 2*\boxOff - 2*\boxWidth, \plotStartY - 2*\arrowLen - 1*\boxHeight - \specOff - 0.5 * \specHeight) {
    \pgfimage[width=\specWidth, height=\specHeight]{images/weak_pipeline/pf_doa_\exNum_\spkNum_weak.pdf} 
};
\draw[\spkColor, thick] 
    (\plotStartX - 2*\boxOff - 2*\boxWidth - 0.5*\specWidth, \plotStartY - 2*\arrowLen - 1*\boxHeight - \specOff - \specHeight)
    rectangle 
    (\plotStartX - 2*\boxOff - 2*\boxWidth + 0.5*\specWidth, \plotStartY - 2*\arrowLen - 1*\boxHeight - \specOff );
\node[anchor=center] at (\plotStartX - 2*\boxOff - 1.5*\boxWidth + \specOff, \plotStartY - 2*\arrowLen - 1*\boxHeight - 0.5*\specHeight - 2*\specOff) {
    \smallFont $\widehat{\theta}_{t}$ 
};

\newcommand\ssfOff{8.3mm}
\draw[line width=\arrowWidth] (\plotStartX - \boxOff, \plotStartY - 3*\arrowLen) rectangle (\plotStartX - \boxOff - \boxWidth, \plotStartY - 3*\arrowLen - \boxHeight);
\node[anchor=center] at (\plotStartX - \boxOff - \ssfOff, \plotStartY - 3*\arrowLen - 0.5 * \boxHeight) {
    \input{images/weak_pipeline/ssf_dnn.tikz}
};
\node[anchor=center] at (\plotStartX - \boxOff - 0.5 * \boxWidth, \plotStartY - 3*\arrowLen - 0.5 * \boxHeight) {
    \largeFont SSF
};

\foreach \x in {1, 0, -1}{
	\ifthenelse{\x < 0}{
    \node[anchor=center] at (\plotStartX + \x * \channelOffset - \boxOff - 0.5 * \boxWidth - \specOff - 0.5 * \specWidth + \channelOffset, \plotStartY - 4*\arrowLen - \boxHeight + \x * \channelOffset) {
    \pgfimage[width=\specWidth, height=\specHeight]{images/self_pipeline/movingdemo_mask_\exNum_\spkNum.pdf}
    };
    \draw[\spkColor, thick] 
    (\plotStartX + \x * \channelOffset - \boxOff - 0.5 * \boxWidth - \specOff +  \channelOffset, \plotStartY - 4*\arrowLen - \boxHeight + \x * \channelOffset - 0.5*\specHeight) 
    rectangle 
    (\plotStartX + \x * \channelOffset - \boxOff - 0.5 * \boxWidth - \specOff - \specWidth + \channelOffset, \plotStartY - 4*\arrowLen - \boxHeight + \x * \channelOffset + 0.5*\specHeight);
    }{}
}

\node[anchor=south west] at (\plotStartX - \boxOff - 0.5 * \boxWidth, \plotStartY - 4*\arrowLen - \boxHeight - \channelOffset) {
    \smallFont $\mathcal{M}_{t}$
};

\foreach \x in {1, 0, -1}{
    \node[anchor=center] at (\plotStartX + \x * \channelOffset, \plotStartY+ \x * \channelOffset + 0.5*\specHeight + \specOff) {
    \pgfimage[width=\specWidth, height=\specHeight]{images/self_pipeline/movingdemo_noisy_\exNum.pdf} 
};
}
\node[anchor=north west] at (\plotStartX, \plotStartY) {
    \smallFont $\mathbf{Y}_{t}$
};

\node[anchor=center] at (\plotStartX- \channelOffset, \plotStartY - 5*\arrowLen - \boxHeight- 0.5 * \specHeight - \specOff - \channelOffset) {
\pgfimage[width=\specWidth, height=\specHeight]{images/self_pipeline/movingdemo_estimate_\exNum_\spkNum.pdf} 
};
\node[anchor=south west] at (\plotStartX- \channelOffset, \plotStartY - 5*\arrowLen - \boxHeight - \channelOffset) {
    \smallFont $\widehat{S}_{t}$
};
\draw[\spkColor, thick] 
    (\plotStartX- \channelOffset - 0.5*\specWidth, \plotStartY - 5*\arrowLen - \boxHeight - \specOff - \channelOffset) 
    rectangle 
    (\plotStartX- \channelOffset + 0.5 * \specWidth, \plotStartY - 5*\arrowLen - \boxHeight- 1 * \specHeight - \specOff - \channelOffset);

\draw[line width=\arrowWidth, ->, >=to] (\plotStartX - 2*\boxOff - 1.5*\boxWidth, \plotStartY - 2*\arrowLen - \boxHeight) -- (\plotStartX - 2*\boxOff - 1.5*\boxWidth, \plotStartY - 3*\arrowLen - 0.5*\boxHeight)--(\plotStartX - \boxOff - \boxWidth, \plotStartY - 3*\arrowLen - 0.5*\boxHeight);
\node[anchor=north east] at (\plotStartX - 3*\boxOff - 2*\boxWidth, \plotStartY) {
    \smallFont $\theta_0$ 
};
\draw[line width=\arrowWidth, ->, >=to] (\plotStartX - 3*\boxOff - 2*\boxWidth, \plotStartY) -- (\plotStartX -3* \boxOff - 2*\boxWidth, \plotStartY - 2*\arrowLen -0.5*\boxHeight) -- (\plotStartX - 2*\boxOff -2*\boxWidth, \plotStartY - 2*\arrowLen - 0.5*\boxHeight);
\node[anchor=south, xshift=\initDoaOff] at (\plotStartX - 2*\boxOff - 2*\boxWidth, \plotStartY) {
    \input{images/weak_pipeline/init_doa.tikz}
};

\end{tikzpicture}

%% file: images/self_pipeline/self_pipeline.tikz
\begin{tikzpicture}

\definecolor{tab_blue}{RGB}{31,119,180}
\definecolor{tab_orange}{RGB}{255,127,14}
\definecolor{tab_green}{RGB}{44,160,44}

\newcommand\plotStartX{0mm}
\newcommand\plotStartY{0mm}

\newcommand\channelOffset{0.5mm}
\newcommand\pfOff{0mm}
\newcommand\specHeight{6mm}
\newcommand\specOff{1mm}
\newcommand\specWidth{8mm}
\newcommand\spkNum{1}
\ifthenelse{\spkNum=0}{\def\spkColor{tab_blue}}{\def\spkColor{tab_orange}}
\newcommand\exNum{3}
\newcommand\initDoaOff{1.5mm}

\newcommand\arrowWidth{0.8pt}
\newcommand\arrowScale{1.5}
\newcommand\arrowR{2mm}
\newcommand\arrowLen{5mm} 
\newcommand\arrowCircleWidth{1.2pt}
\newcommand\crossR{1mm}
\pgfmathsetmacro{\crossOff}{\crossR * cos(45)}
\newcommand\zWidth{5mm}

\newcommand\largeFont{\footnotesize}
\newcommand\smallFont{\scriptsize}

\newcommand\boxHeight{5mm}
\newcommand\boxWidth{10mm}
\newcommand\boxCorner{4pt}
\newcommand\boxOff{3mm}

\pgfmathsetmacro{\recWidth}{
    4*\boxOff + 2*\boxWidth
}
\foreach \x in {1, 0, -1} {
    \pgfmathsetmacro{\absX}{abs(\x)}

    \ifthenelse{\x > 0}{\def\theColor{gray!40}}{}
    \ifthenelse{\x < 0}{\def\theColor{black}}{}
    \ifthenelse{\x = 0}{\def\theColor{gray!70}}{}
    
    \draw[line width=\arrowWidth, color=\theColor, ->, >=to] (\plotStartX + \x * \channelOffset, \plotStartY+ \x * \channelOffset) -- (\plotStartX + \x * \channelOffset, \plotStartY - 3*\arrowLen - \boxHeight + \crossR+ \x * \channelOffset);

    \draw[line width=\arrowWidth, color=\theColor, ->, >=to] (\plotStartX + \x * \channelOffset, \plotStartY - \arrowLen+ \x * \channelOffset) -- (\plotStartX + \x * \channelOffset + \boxOff + 0.5 * \boxWidth, \plotStartY - \arrowLen+ \x * \channelOffset) -- (\plotStartX + \x * \channelOffset + \boxOff + 0.5 * \boxWidth, \plotStartY - 2*\arrowLen);
    \fill[\theColor] (\plotStartX + \x * \channelOffset, \plotStartY - \arrowLen+ \x * \channelOffset) circle (\arrowCircleWidth);

    \draw[line width=\arrowWidth, color=\theColor, ->, >=to] (\plotStartX + \x * \channelOffset + \boxOff + 0.5 * \boxWidth, \plotStartY - 2*\arrowLen - \boxHeight) -- (\plotStartX + \x * \channelOffset + \boxOff + 0.5 * \boxWidth, \plotStartY - 3*\arrowLen - \boxHeight + \x * \channelOffset) -- (\plotStartX + \x * \channelOffset + \crossR, \plotStartY - 3*\arrowLen - \boxHeight + \x * \channelOffset);

    \ifthenelse{\x < 0}{
        \draw[line width=\arrowWidth, color=\theColor, ->, >=to] (\plotStartX + \x * \channelOffset, \plotStartY - 3*\arrowLen - \boxHeight - \crossR + \x * \channelOffset) -- (\plotStartX + \x * \channelOffset, \plotStartY - 5*\arrowLen - \boxHeight+ \x * \channelOffset);
        \fill[\theColor] (\plotStartX + \x * \channelOffset, \plotStartY - 4*\arrowLen - \boxHeight + \x * \channelOffset) circle (\arrowCircleWidth);
    }{
        \draw[line width=\arrowWidth, color=\theColor] (\plotStartX + \x * \channelOffset, \plotStartY - 3*\arrowLen - \boxHeight - \crossR + \x * \channelOffset) -- (\plotStartX + \x * \channelOffset, \plotStartY - 4*\arrowLen - \boxHeight + \x * \channelOffset - 0.5*\arrowWidth);
    }

    \draw[line width=\arrowWidth, color=\theColor, ->, >=to] (\plotStartX + \x * \channelOffset, \plotStartY - 4*\arrowLen - \boxHeight + \x * \channelOffset) -- (\plotStartX + \recWidth - 0.5 * \zWidth + \x * \channelOffset - \arrowWidth, \plotStartY - 4*\arrowLen - \boxHeight + \x * \channelOffset);
    \draw[line width=\arrowWidth, color=\theColor, ->, >=to] (\plotStartX + \x * \channelOffset + \recWidth, \plotStartY - 4*\arrowLen - \boxHeight + \x * \channelOffset- 0.5 * \zWidth) -- (\plotStartX + \x * \channelOffset + \recWidth, \plotStartY  + \boxHeight+ \x * \channelOffset+ \pfOff - \arrowLen) -- (\plotStartX + \x * \channelOffset + \recWidth - \boxOff - 0.5*\boxWidth, \plotStartY + \boxHeight+ \x * \channelOffset + \pfOff - \arrowLen) -- (\plotStartX + \x * \channelOffset + \recWidth - \boxOff - 0.5*\boxWidth, \plotStartY - 2*\arrowLen + \boxHeight + \pfOff);

    \fill[color=white] (\plotStartX+ \x * \channelOffset, \plotStartY - 3*\arrowLen - \boxHeight+ \x * \channelOffset) circle (\crossR);
    \draw[color=\theColor] (\plotStartX+ \x * \channelOffset, \plotStartY - 3*\arrowLen - \boxHeight+ \x * \channelOffset) circle (\crossR);
    \draw[color=\theColor] (\plotStartX + \crossOff + \x * \channelOffset, \plotStartY - 3*\arrowLen - \boxHeight+ \crossOff+ \x * \channelOffset) -- (\plotStartX - \crossOff+ \x * \channelOffset, \plotStartY - 3*\arrowLen - \boxHeight- \crossOff+ \x * \channelOffset);
    \draw[color=\theColor] (\plotStartX - \crossOff+ \x * \channelOffset, \plotStartY - 3*\arrowLen - \boxHeight+\crossOff+ \x * \channelOffset) -- (\plotStartX  +\crossOff+ \x * \channelOffset, \plotStartY - 3*\arrowLen - \boxHeight -\crossOff+ \x * \channelOffset);
    
    \draw[line width=2*\arrowWidth, anchor=mid, color=\theColor] (\plotStartX + \recWidth - 0.5 * \zWidth + \x * \channelOffset, \plotStartY - 4*\arrowLen - \boxHeight - 0.5 * \zWidth + \x * \channelOffset) rectangle (\plotStartX + \recWidth + 0.5 * \zWidth + \x * \channelOffset, \plotStartY - 4*\arrowLen - \boxHeight + 0.5 * \zWidth + \x * \channelOffset);
    \fill[color=white, anchor=mid] (\plotStartX + \recWidth - 0.5 * \zWidth + \x * \channelOffset, \plotStartY - 4*\arrowLen - \boxHeight  - 0.5 * \zWidth + \x * \channelOffset) rectangle (\plotStartX + \recWidth + 0.5 * \zWidth + \x * \channelOffset, \plotStartY - 4*\arrowLen - \boxHeight  + 0.5 * \zWidth + \x * \channelOffset);
    \ifthenelse{\x < 0}{
        \node at (\plotStartX  + \recWidth +\x * \channelOffset, \plotStartY - 4*\arrowLen - \boxHeight+ \x * \channelOffset) {\scriptsize $z^{\scriptscriptstyle{-1}}$};
    }{}
    
}

\pgfmathsetlengthmacro{\w}{\boxWidth-\arrowWidth}
\pgfmathsetlengthmacro{\h}{\boxHeight-\arrowWidth}
\node[anchor=center] at (\plotStartX + \recWidth - \boxOff - 0.5*\boxWidth, \plotStartY - 2*\arrowLen + 0.5*\boxHeight + \pfOff) {
    \pgfimage[width=\w, height=\h]{images/self_pipeline/particles_\exNum_\spkNum.png} 
};
\draw[line width=\arrowWidth] (\plotStartX + \recWidth - \boxOff - \boxWidth, \plotStartY + \boxHeight + \pfOff - 2*\arrowLen) rectangle (\plotStartX + \recWidth - \boxOff, \plotStartY + \pfOff - 2*\arrowLen);
\node[anchor=center] at (\plotStartX + \recWidth - \boxOff - 0.5*\boxWidth, \plotStartY - 2*\arrowLen + 0.5*\boxHeight + \pfOff) {
    \largeFont PF
};
\node[anchor=north west] at (\plotStartX + \recWidth - \boxOff - 0.5*\boxWidth, \plotStartY + \boxHeight+ \pfOff - \arrowLen) {
     \smallFont $\mathbf{\widehat{S}}_{t-1}$
};
\node[anchor=center] at (\plotStartX + \recWidth - \boxOff - 0.5*\boxWidth, \plotStartY - 2*\arrowLen - 0.5*\boxHeight - \specOff - 0.5 * \specHeight) {
    \pgfimage[width=\specWidth, height=\specHeight]{images/self_pipeline/pf_doa_\exNum_\spkNum.pdf} 
};
\draw[\spkColor, thick] 
    (\plotStartX + \recWidth - \boxOff - 0.5*\boxWidth + 0.5 * \specWidth, \plotStartY - 2*\arrowLen - 0.5*\boxHeight - \specOff - \specHeight) 
    rectangle 
    (\plotStartX + \recWidth - \boxOff - 0.5*\boxWidth - 0.5 * \specWidth, \plotStartY - 2*\arrowLen - 0.5*\boxHeight - \specOff);
\node[anchor=south west] at (\plotStartX + \boxOff + \boxWidth, \plotStartY - 2*\arrowLen - 0.5 * \boxHeight - 0.5*\specOff) {
    \smallFont $\widehat{\theta}_{t-1}$ 
};

\newcommand\ssfOff{1.65mm}
\draw[line width=\arrowWidth] (\plotStartX + \boxOff, \plotStartY - 2*\arrowLen) rectangle (\plotStartX + \boxOff + \boxWidth, \plotStartY - 2*\arrowLen - \boxHeight);
\node[anchor=center] at (\plotStartX + \boxOff + \ssfOff, \plotStartY - 2*\arrowLen - 0.5 * \boxHeight) {
    \input{images/weak_pipeline/ssf_dnn.tikz}
};
\node[anchor=center] at (\plotStartX + \boxOff + 0.5 * \boxWidth, \plotStartY - 2*\arrowLen - 0.5 * \boxHeight) {
    \largeFont SSF
};

\foreach \x in {1, 0, -1}{
    \node[anchor=center] at (\plotStartX + \x * \channelOffset + \boxOff + 0.5 * \boxWidth + \specOff + 0.5 * \specWidth, \plotStartY - 3*\arrowLen - \boxHeight + \x * \channelOffset) {
    \pgfimage[width=\specWidth, height=\specHeight]{images/self_pipeline/movingdemo_mask_\exNum_\spkNum.pdf}
    };
    \draw[\spkColor, thick] 
    (\plotStartX + \x * \channelOffset + \boxOff + 0.5 * \boxWidth + \specOff, \plotStartY - 3*\arrowLen - \boxHeight + \x * \channelOffset - 0.5*\specHeight) 
    rectangle 
    (\plotStartX + \x * \channelOffset + \boxOff + 0.5 * \boxWidth + \specOff + \specWidth, \plotStartY - 3*\arrowLen - \boxHeight + \x * \channelOffset + 0.5*\specHeight);

}
\node[anchor=south east] at (\plotStartX + \boxOff + 0.5 * \boxWidth, \plotStartY - 3*\arrowLen - \boxHeight) {
    \smallFont $\mathbfcal{M}_{t}$
};

\foreach \x in {1, 0, -1}{
    \node[anchor=center] at (\plotStartX + \x * \channelOffset, \plotStartY+ \x * \channelOffset + 0.5*\specHeight + \specOff) {
    \pgfimage[width=\specWidth, height=\specHeight]{images/self_pipeline/movingdemo_noisy_\exNum.pdf} 
};
}
\node[anchor=north west] at (\plotStartX, \plotStartY) {
    \smallFont $\mathbf{Y}_{t}$
};

\node[anchor=center] at (\plotStartX- \channelOffset, \plotStartY - 5*\arrowLen - \boxHeight- 0.5 * \specHeight - \specOff - \channelOffset) {
\pgfimage[width=\specWidth, height=\specHeight]{images/self_pipeline/movingdemo_estimate_\exNum_\spkNum.pdf} 
};
\node[anchor=south west] at (\plotStartX- \channelOffset, \plotStartY - 5*\arrowLen - \boxHeight - \channelOffset) {
    \smallFont $\widehat{S}_{t}$
};
\draw[\spkColor, thick] 
    (\plotStartX- \channelOffset - 0.5*\specWidth, \plotStartY - 5*\arrowLen - \boxHeight - \specOff - \channelOffset) 
    rectangle 
    (\plotStartX- \channelOffset + 0.5 * \specWidth, \plotStartY - 5*\arrowLen - \boxHeight- 1 * \specHeight - \specOff - \channelOffset);

\newcommand\swArrowLen{\arrowLen}
\newcommand\swLongLen{3mm}
\newcommand\swShortLen{2mm}
\newcommand\swArrowR{1.5mm}
\newcommand\swArrowWidth{0.4pt}
\pgfmathsetmacro{\swLongOff}{\swLongLen * cos(45)}
\pgfmathsetmacro{\swShortOff}{\swShortLen * cos(45)}
\definecolor{sw_color}{RGB}{255,0,0}
\draw[line width=\arrowWidth, ->, >=to] (\plotStartX + \recWidth - \boxOff - 0.5*\boxWidth, \plotStartY + \pfOff - 2*\arrowLen) -- (\plotStartX + \recWidth - \boxOff - 0.5*\boxWidth, \plotStartY - 2*\arrowLen - 0.5*\boxHeight)--(\plotStartX + \boxOff + \boxWidth, \plotStartY - 2*\arrowLen - 0.5*\boxHeight);
\node[anchor=north east] at (\plotStartX + 2*\boxOff + \boxWidth, \plotStartY) {
    \smallFont $\theta_0$ 
};
\draw[line width=\arrowWidth, ->, >=to] (\plotStartX + 2*\boxOff + \boxWidth, \plotStartY) -- (\plotStartX + 2*\boxOff + \boxWidth, \plotStartY - 2*\arrowLen + 0.5*\boxHeight + \pfOff) -- (\plotStartX + 3*\boxOff + \boxWidth, \plotStartY - 2*\arrowLen + 0.5*\boxHeight + \pfOff);
\node[anchor=south, xshift=\initDoaOff] at (\plotStartX + 2*\boxOff + \boxWidth, \plotStartY) {
    \input{images/weak_pipeline/init_doa.tikz}
};

\end{tikzpicture}

%% file: images/trajectories/trajectories.tikz
\begin{tikzpicture}
    
\newcommand\velWidth{1.8cm}
\newcommand\velHeight{2.525cm}
\newcommand\velOff{3.13mm}
\newcommand\doaWidth{1.5cm}
\newcommand\doaHeight{8mm}
\newcommand\doaDist{2mm}
\newcommand\doaLabelDist{3mm}
\newcommand\lrDist{10mm}
\newcommand\tbDist{8mm}
\newcommand\plotStartX{0mm}
\newcommand\plotStartY{0mm}
\newcommand\majorTick{0.75mm}
\newcommand\minorTick{0.5mm}
\newcommand\tickSize{7pt}
\newcommand\tickSkip{9pt}
\newcommand\timeDist{3.1mm} 
\newcommand\legendTick{4mm}
\newcommand\legendTickTop{-0.25mm}
\newcommand\legendTickBottom{-0.5mm}
\pgfmathsetlengthmacro{\legendTickMiddle}{0.5 * \legendTickTop + 0.5 * \legendTickBottom}
\newcommand\legendOff{1mm}
\newcommand\legendYOff{-2mm} 
\newcommand\centerOff{-1.1cm}
\newcommand\legendWidth{4cm}
\newcommand\legendSkip{4mm}
\newcommand\legendOffProp{2.7cm}
\newcommand\labelWidth{1pt}
\newcommand\descriptX{2.7mm}
\newcommand\descriptY{2.2mm}
\newcommand\legendXBoxOff{1.25cm}
\newcommand\legendYBoxOff{4mm}
\newcommand\legendXBoxWidth{2.8cm}
\newcommand\legendYBoxHeight{2.05cm}

\pgfmathsetlengthmacro{\velLegendXOff}{- 12.5*\majorTick}
\newcommand\velLegendYOff{0.5mm} 
\newcommand\velLegendXBoxOff{-1.1cm}
\newcommand\velLegendYBoxOff{1.25mm}
\newcommand\velLegendYBoxHeight{4mm}
\newcommand\velLegendXBoxWidth{2.9cm}

\node[anchor=center] at (\plotStartX + 0.5 * \velWidth, \plotStartY + 0.5 * \velHeight) {%
    \pgfimage[height=\velHeight, width=\velWidth]{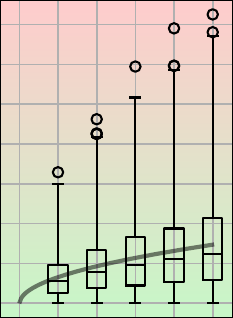} 
};
\pgfmathsetmacro{\timeOff}{\doaWidth / 5}
\foreach \t in {0, ..., 5}{
    \pgfmathtruncatemacro\tmp{int(\t/2) * 2}
    \pgfmathsetmacro{\time}{int(\t)}
    \ifnum\tmp=\t
        \pgfmathsetmacro{\tickLen}{\majorTick}
        \node[anchor=north] at (\plotStartX + \t * \timeOff + 0.5 * \timeOff, \plotStartY) {\fontsize{\tickSize}{\tickSkip}\selectfont \time};
    \else
        \pgfmathsetmacro{\tickLen}{\minorTick}
    \fi
    \draw[line width=\tickWidth] (\plotStartX + \t * \timeOff + 0.5 * \timeOff, \plotStartY) --  (\plotStartX + \t * \timeOff + 0.5 * \timeOff, \plotStartY - \tickLen);
}
\foreach \v in {0, ..., 7}{
    \pgfmathtruncatemacro\tmp{int(\v/2) * 2}
    \pgfmathsetmacro{\vel}{int(\v)}
    \ifnum\tmp=\v
        \pgfmathsetmacro{\tickLen}{\majorTick}
        \node[anchor=east] at (\plotStartX, \plotStartY+ 0.5 * \timeOff + \v * \velOff) {\fontsize{\tickSize}{\tickSkip}\selectfont \vel};
    \else
        \pgfmathsetmacro{\tickLen}{\minorTick}
    \fi
    \draw[line width=\tickWidth] (\plotStartX, \plotStartY+ 0.45 * \timeOff + \v * \velOff) --  (\plotStartX - \tickLen, \plotStartY  + 0.45 * \timeOff + \v * \velOff);
}
\begin{scope}[>=Latex]
\draw[|->, line width=\tickWidth]    (\plotStartX - 5*\majorTick,\plotStartY+ 0.5 * \timeOff) -- (\plotStartX - 5*\majorTick,\plotStartY+ 0.5 * \timeOff + 3 * \velOff);
\draw[|->, line width=\tickWidth]    (\plotStartX - 5*\majorTick,\plotStartY+ 0.5 * \timeOff + 3 * \velOff) -- (\plotStartX - 5*\majorTick,\plotStartY+ \velHeight);
\end{scope}
\node[anchor=south, rotate=90] at (\plotStartX - 5*\majorTick, \plotStartY+ 0.5 * \timeOff + 1.5 * \velOff) {\footnotesize walking};
\node[anchor=south, rotate=90] at (\plotStartX - 5*\majorTick, \plotStartY+ 1.5 * \velOff + 0.5*\velHeight) {\footnotesize running};
\node[anchor=south, rotate=90] at (\plotStartX - 9*\majorTick, \plotStartY + 0.5*\velHeight) {\footnotesize abs. velocity [{\scriptsize $\frac{\mathrm{m}}{\mathrm{s}}$}] };

\draw[color=black, line width=\labelWidth, opacity=0.5] (\plotStartX + \velLegendXOff, \plotStartY + 1*\doaHeight + \tbDist + \doaWidth+\legendTickBottom - 0.1 * \legendSkip + \legendYOff+ \velLegendYOff) --  (\plotStartX  + \velLegendXOff + \legendTick, \plotStartY + 1*\doaHeight + \tbDist + \doaWidth+ \legendTickBottom - 0.1 * \legendSkip + \legendYOff + \velLegendYOff);
\node[anchor=west, yshift=-1mm] at (\plotStartX  + \velLegendXOff + \legendOff + \legendTick, \plotStartY + 1*\doaHeight + \tbDist + \doaWidth  + \legendYOff + \velLegendYOff) {%
        \footnotesize exp.\,value\,$\mathbb{E}\hspace*{-1pt}\{ |v_t| \}$
    };
\draw[draw, rounded corners=\rectEdge, line width=\tickWidth]
            (\plotStartX + \velLegendXBoxOff, \plotStartY + \velLegendYBoxOff+ \velHeight) rectangle (\plotStartX + \velLegendXBoxOff + \velLegendXBoxWidth, \plotStartY + \velLegendYBoxOff + \velLegendYBoxHeight+ \velHeight);

\foreach \x / \w / \h / \l / \i [
    remember=\dl as \lastLen (initially 0), 
    evaluate=\l as \dl using \l+\lastLen,
]in {
    0/7.8/7.8/5/0, 1/7.65/7.6/5/26, 2/6.4/7.4/5/17, 3/7.9/7.3/5/33, 4/8.0/6.6/5/14, 5/7.6/7.4/5/4, 6/5.1/5.4/5/42
    }{
    
    \pgfmathsetmacro{\filenum}{int(\i)}
    \def\doaname{pf_doa_\filenum}
    \def\roomname{room_\filenum}
    
    \pgfmathsetlengthmacro{\dWidth}{\l / 5 * \doaWidth}
    \pgfmathsetlengthmacro{\dOff}{\lastLen / 5 * \doaWidth}
    \pgfmathsetmacro{\dOffVal}{int(abs(\dOff))}
    \pgfmathsetmacro{\dMaxVal}{6 * \doaWidth}
    \ifthenelse{\dOffVal > \dMaxVal}{}{ 
    \node[anchor=mid] at (\plotStartX + \velWidth + \lrDist + \dOff + \x * \doaDist + 0.5 * \dWidth, \plotStartY + 0.5 \doaHeight) {%
        \pgfimage[height=\doaHeight, width=\dWidth]{images/trajectories/pf/\doaname.pdf} 
    };
    \pgfmathsetmacro{\timeNum}{int(\l)}
    \pgfmathsetmacro{\timeOff}{\doaWidth / 5}
    \pgfmathsetmacro{\doaOff}{\doaHeight / 4 - 0.025mm}
    \foreach \t in {0, ..., \timeNum}{
        \pgfmathtruncatemacro\tmp{int(\t/2) * 2}
        \ifnum\tmp=\t
            \pgfmathsetmacro{\tickLen}{\majorTick}
            \pgfmathsetmacro{\time}{int(\t)}
            \node[anchor=north] at (\plotStartX + \velWidth + \lrDist + \dOff + \x * \doaDist + \t * \timeOff, \plotStartY) {\fontsize{\tickSize}{\tickSkip}\selectfont \time};
        \else
            \pgfmathsetmacro{\tickLen}{\minorTick}
        \fi
        \draw[line width=\tickWidth] (\plotStartX + \velWidth + \lrDist + \dOff + \x * \doaDist + \t * \timeOff, \plotStartY) --  (\plotStartX + \velWidth + \lrDist + \dOff + \x * \doaDist + \t * \timeOff, \plotStartY - \tickLen);
    }
    \foreach \d in {0, ..., 4}{
        \pgfmathtruncatemacro\tmp{int(\d/2) * 2}
        \ifnum\tmp=\d
            \pgfmathsetmacro{\tickLen}{\majorTick}
            \ifthenelse{\x = 0}{
                \pgfmathsetmacro{\doa}{int(\d * 90 - 180)}
                \node[anchor=east] at (\plotStartX + \velWidth + \lrDist + \dOff + \x * \doaDist ,\plotStartY + \d * \doaOff) {\fontsize{\tickSize}{\tickSkip}\selectfont \doa};
            }{}
        \else
            \pgfmathsetmacro{\tickLen}{\minorTick}
        \fi
        \draw[line width=\tickWidth] (\plotStartX + \velWidth + \lrDist + \dOff + \x * \doaDist - \tickLen,\plotStartY + \d * \doaOff) --  (\plotStartX + \velWidth + \lrDist + \dOff + \x * \doaDist,\plotStartY + \d * \doaOff);
    }
    
    \pgfmathsetlengthmacro{\W}{\w / 8 * \doaWidth}
    \pgfmathsetlengthmacro{\H}{\h / 8 * \doaWidth}
    \node[anchor=center] at (\plotStartX + \velWidth + \lrDist + \dOff + \x * \doaDist + 0.5 * \W, \plotStartY + 1*\doaHeight + \tbDist + 0.5 * \H) {%
        \pgfimage[height=\H, width=\W]{images/trajectories/room/\roomname.pdf} 
    };
    \pgfmathsetmacro{\widthNum}{int(\w)}
    \pgfmathsetlengthmacro{\widthOff}{\doaWidth / 8}
    \foreach \wn in {0, ..., \widthNum}{
        \pgfmathtruncatemacro\tmp{int(\wn/3) * 3}
        \ifnum\tmp=\wn
            \pgfmathsetmacro{\tickLen}{\majorTick}  
            \node[anchor=north] at ((\plotStartX + \velWidth + \lrDist + \dOff + \x * \doaDist + \wn * \widthOff, \plotStartY + \doaHeight + \tbDist) {\fontsize{\tickSize}{\tickSkip}\selectfont \wn};
        \else
            \pgfmathsetmacro{\tickLen}{\minorTick}
        \fi
        \draw[line width=\tickWidth] (\plotStartX + \velWidth + \lrDist + \dOff + \x * \doaDist + \wn * \widthOff, \plotStartY + \doaHeight + \tbDist) --  (\plotStartX + \velWidth + \lrDist + \dOff + \x * \doaDist + \wn * \widthOff, \plotStartY - \tickLen + \doaHeight + \tbDist);
    }
    \pgfmathsetmacro{\heightNum}{int(\h)}
    \pgfmathsetlengthmacro{\heightOff}{\doaWidth / 8}
    \foreach \hn in {0, ..., \heightNum}{
        \pgfmathtruncatemacro\tmp{int(\hn/3) * 3}
        \ifnum\tmp=\hn
            \pgfmathsetmacro{\tickLen}{\majorTick}
            \ifthenelse{\x = 0}{
                \node[anchor=east] at (\plotStartX + \velWidth + \lrDist + \dOff + \x * \doaDist, \plotStartY + \doaHeight + \tbDist + \hn * \heightOff) {\fontsize{\tickSize}{\tickSkip}\selectfont \hn};
            }{}
        \else
            \pgfmathsetmacro{\tickLen}{\minorTick}
        \fi
        \draw[line width=\tickWidth] (\plotStartX + \velWidth + \lrDist + \dOff + \x * \doaDist - \tickLen, \plotStartY + \doaHeight + \tbDist + \hn * \heightOff) --  (\plotStartX + \velWidth + \lrDist + \dOff + \x * \doaDist, \plotStartY + \doaHeight + \tbDist + \hn * \heightOff);
    }
    }
    
}

\node[anchor=north] at (\plotStartX + 0.5 * \velWidth, \plotStartY- 3*\majorTick) {\footnotesize time [s]};
\node[anchor=north] at (0.5 * \textwidth + \centerOff, \plotStartY- 3*\majorTick) {\footnotesize time [s]};
\node[anchor=north] at (0.5 * \textwidth + \centerOff, \plotStartY- 3*\majorTick + \doaHeight + \tbDist) {\footnotesize room width [m]};
\node[anchor=south, rotate=90] at (\plotStartX + \velWidth + \lrDist - 4.5mm, \plotStartY + 0.5 * \doaHeight) {\footnotesize DoA [°]};
\node[anchor=south, rotate=90] at (\plotStartX + \velWidth + \lrDist - 2mm, \plotStartY + \doaHeight+ 0.5 * \doaWidth + \tbDist) {\footnotesize length [m]};

\draw[draw, rounded corners=\rectEdge, line width=\tickWidth]
            (\textwidth - \legendXBoxOff, \plotStartY + \doaHeight + \legendYBoxOff + \legendYOff) rectangle (\textwidth - \legendXBoxOff - \legendXBoxWidth, \plotStartY + \doaHeight + \legendYBoxOff + \legendYBoxHeight + \legendYOff);

\newcounter{mycounter}
\setcounter{mycounter}{0}
\foreach \label in {mic. array, start\,/\,stop, oracle, PF-CV\,concat., PF-CV\,AR\,(ours)}{
    \node[anchor=west, yshift=-1mm] at (\textwidth - \legendWidth + 2* \legendOff + \legendTick, \plotStartY + 1*\doaHeight + \tbDist + \doaWidth - \themycounter * \legendSkip + \legendYOff) {%
        \footnotesize \label
    };
    \stepcounter{mycounter}
}
\node[isosceles triangle, isosceles triangle apex angle=60,
    draw,
    fill=black,
    minimum size =0.5mm, rotate=-30, scale=0.25] (T60)at (\textwidth - \legendWidth + \legendOff + 0.5 * \legendTick, \plotStartY + 1*\doaHeight + \tbDist + \doaWidth + \legendTickBottom - 0.1 * \legendSkip + \legendYOff){};

\node[] at (\textwidth - \legendWidth + \legendOff + 0.5 * \legendTick, \plotStartY + 1*\doaHeight + \tbDist + \doaWidth + \legendTickBottom - 1.1 * \legendSkip + \legendYOff) {
	\footnotesize /
};
\draw (\textwidth - \legendWidth + \legendOff + \legendTick - 1pt, \plotStartY + 1*\doaHeight + \tbDist + \doaWidth + \legendTickTop- 1 * \legendSkip + \legendYOff) node[cross=2pt, color=tab_blue, line width=0.5*\labelWidth, anchor=center]{};
\draw (\textwidth - \legendWidth + \legendOff + \legendTick - 1pt, \plotStartY + 1*\doaHeight + \tbDist + \doaWidth - 0.25*\legendSkip- 1 * \legendSkip + \legendTickBottom + \legendYOff) node[cross=2pt, color=tab_orange, line width=0.5*\labelWidth, anchor=center]{};
\fill[tab_blue] (\textwidth - \legendWidth + \legendOff + 0.75pt, \plotStartY + 1*\doaHeight + \tbDist + \doaWidth + \legendTickTop- 1 * \legendSkip + \legendYOff) circle (1.5pt);
\fill[tab_orange] (\textwidth - \legendWidth + \legendOff + 0.75pt, \plotStartY + 1*\doaHeight + \tbDist + \doaWidth - 0.25*\legendSkip- 1 * \legendSkip + \legendTickBottom + \legendYOff) circle (1.5pt);

\draw[color=tab_blue, line width=2*\labelWidth, opacity=0.5] (\textwidth - \legendWidth + \legendOff, \plotStartY + 1*\doaHeight + \tbDist + \doaWidth- 2 * \legendSkip + \legendTickTop + \legendYOff) --  (\textwidth - \legendWidth + \legendOff + \legendTick, \plotStartY + 1*\doaHeight + \tbDist + \doaWidth- 2 * \legendSkip + \legendTickTop + \legendYOff);
\draw[color=tab_orange, line width=2*\labelWidth, opacity=0.5] (\textwidth - \legendWidth + \legendOff, \plotStartY + 1*\doaHeight + \tbDist + \doaWidth - 0.25*\legendSkip- 2 * \legendSkip + \legendTickBottom + \legendYOff) --  (\textwidth - \legendWidth + \legendOff + \legendTick, \plotStartY + 1*\doaHeight + \tbDist + \doaWidth - 0.25*\legendSkip- 2 * \legendSkip + \legendTickBottom + \legendYOff);

 \newcommand\pfYoff{-0.2mm}
\draw[color=tab_blue, line width=\labelWidth, dash pattern=on 1pt off 1pt] (\textwidth - \legendWidth + \legendOff, \plotStartY + 1*\doaHeight + \tbDist + \doaWidth- 3 * \legendSkip + \legendTickTop + \legendYOff + \pfYoff) --  (\textwidth - \legendWidth + \legendOff + \legendTick, \plotStartY + 1*\doaHeight + \tbDist + \doaWidth- 3 * \legendSkip + \legendTickTop + \legendYOff + \pfYoff);
\draw[color=tab_orange, line width=\labelWidth, dash pattern=on 1pt off 1pt] (\textwidth - \legendWidth + \legendOff, \plotStartY + 1*\doaHeight + \tbDist + \doaWidth - 0.25*\legendSkip- 3 * \legendSkip + \legendTickBottom + \legendYOff + \pfYoff) --  (\textwidth - \legendWidth + \legendOff + \legendTick, \plotStartY + 1*\doaHeight + \tbDist + \doaWidth - 0.25*\legendSkip- 3 * \legendSkip + \legendTickBottom + \legendYOff + \pfYoff);

\draw[color=tab_blue, line width=\labelWidth, dash pattern=on 3pt off 1pt] (\textwidth - \legendWidth + \legendOff, \plotStartY + 1*\doaHeight + \tbDist + \doaWidth- 4 * \legendSkip + \legendTickTop + \legendYOff) --  (\textwidth - \legendWidth + \legendOff + \legendTick, \plotStartY + 1*\doaHeight + \tbDist + \doaWidth- 4 * \legendSkip + \legendTickTop + \legendYOff);
\draw[color=tab_orange, line width=\labelWidth, dash pattern=on 3pt off 1pt] (\textwidth - \legendWidth + \legendOff, \plotStartY + 1*\doaHeight + \tbDist + \doaWidth - 0.25*\legendSkip- 4 * \legendSkip + \legendTickBottom + \legendYOff) --  (\textwidth - \legendWidth + \legendOff + \legendTick, \plotStartY + 1*\doaHeight + \tbDist + \doaWidth - 0.25*\legendSkip- 4 * \legendSkip + \legendTickBottom + \legendYOff);

\end{tikzpicture}

%% file: images/tracking_enhancement_correlation/tracking_enhancement_correlation.tikz
\begin{tikzpicture}

\newcommand\plotStartX{0mm}
\newcommand\plotStartY{0mm}
\newcommand\plotHeight{2.5cm}
\newcommand\plotWidth{3.8cm}
\newcommand\plotOff{3mm}
\newcommand\majorTick{0.75mm}
\newcommand\minorTick{0.5mm}
\newcommand\tickXDist{6.75mm}
\newcommand\tickXOffY{-6mm}
\newcommand\tickXOffX{-4mm}
\newcommand\tickRotation{45}
\newcommand\legendXOff{1mm}
\newcommand\legendXMarkerStart{0mm}
\newcommand\legendXStart{0mm}
\newcommand\legendHeight{3.5mm}
\newcommand\legendYOff{1mm}
\newcommand\legendYOffText{1.75mm}

\foreach \i in {0, 1}{
    
    \node[anchor=center] at (\plotStartX + 0.5 * \plotWidth + \i * \plotWidth + \i * \plotOff, \plotStartY + 0.5 * \plotHeight) {%
        \pgfimage[height=\plotHeight, width=\plotWidth]{images/tracking_enhancement_correlation/delta_sisdr_\i.png} 
    };

   \pgfmathsetlengthmacro{\tickYDist}{\plotHeight / 8}
    \foreach \y in {0, ..., 8}{
        \pgfmathtruncatemacro\tmp{int(\y/2) * 2}
        \pgfmathsetmacro{\yticklabel}{int(\y * 2 + 2)}
        \ifnum\tmp=\y
            \pgfmathsetmacro{\tickLen}{\majorTick}
            \ifthenelse{\i = 0}{
                \node[anchor=east] at (\plotStartX+ \i * \plotWidth + \i * \plotOff, \plotStartY  + \y * \tickYDist) {\footnotesize \yticklabel};
            }{}
        \else
            \pgfmathsetmacro{\tickLen}{\minorTick}
        \fi
        \draw[line width=\tickWidth] (\plotStartX+ \i * \plotWidth + \i * \plotOff, \plotStartY  + \y * \tickYDist) --  (\plotStartX - \tickLen+ \i * \plotWidth + \i * \plotOff, \plotStartY  + \y * \tickYDist);
    }
    \ifthenelse{\i = 0}{
        \node[anchor=center, rotate=90, yshift=7mm] at (\plotStartX+ \i * \plotWidth + \i * \plotOff, \plotStartY  + 0.5 * \plotHeight) {\footnotesize $\Delta$\acs{sisdr}\,[dB]\,$\rightarrow$};
    }{}
    \foreach \exp / \tick in {
        -1/7, -1/8, -1/9,
        0/1, 0/2, 0/3, 0/4, 0/5, 0/6, 0/7, 0/8, 0/9,
        1/1, 1/2, 1/3, 1/4, 1/5, 1/6, 1/7, 1/8, 1/9,
        2/1, 2/2
    }{
        \pgfmathsetmacro{\tickXMulti}{
            (log10(\tick) + \exp + 1 - log10(7)) / (2 + log10(2) + 1 - log10(7))
        }
        \pgfmathsetlengthmacro{\tickXOff}{
          \tickXMulti * \plotWidth
        }
        \ifthenelse{\tick = 1}{
            \def\tickLen{\majorTick}
            \pgfmathsetmacro{\xticklabel}{int(10^\exp)}
            \node[anchor=north] at (\plotStartX + \i * \plotWidth + \i * \plotOff + \tickXOff, \plotStartY) {\footnotesize \xticklabel};
        }{
            \def\tickLen{\minorTick}
        }
        \draw[line width=\tickWidth] (\plotStartX + \i * \plotWidth + \i * \plotOff + \tickXOff, \plotStartY) --  (\plotStartX + \i * \plotWidth + \i * \plotOff + \tickXOff, \plotStartY - \tickLen);
    }
    \node[anchor=center] at (\plotStartX + 0.5 * \plotWidth + \i * \plotWidth + \i * \plotOff, \plotStartY + \tickXOffY) {%
        \footnotesize \acs{mae}\,[°]\,$\leftarrow$
        };

    \draw[draw, rounded corners=\rectEdge, line width=\tickWidth]
            (\plotStartX + \i * \plotWidth + \i * \plotOff, \plotStartY+ \plotHeight+ \legendYOff) rectangle (\plotStartX + \plotWidth+ \i * \plotWidth + \i * \plotOff, \plotStartY+ \plotHeight+ \legendYOff + \legendHeight);
    \ifthenelse{\i = 0}{
        \def\pfVersion{{\acs{pf}-\acs{rw}}}
    }{
        \def\pfVersion{{\acs{pf}-\acs{cv}}}
    }
    \node[anchor=mid west] at (\plotStartX + \i * \plotWidth + \i * \plotOff + \legendXStart + \legendXOff, \plotStartY+ \plotHeight+ \legendYOff+ \legendYOffText) {%
        \footnotesize \pfVersion
        };
    \foreach \l / \pipelineVersion in {
        1/{concat.}, 2/{\acs{ar}\,(ours)}
    }{
        \ifthenelse{\l = 1}{
            \def\origColor{orig_green}
            \def\customColor{custom_green}
        }{
            \def\origColor{orig_purple}
            \def\customColor{custom_purple}
        }
        \ifthenelse{\i = 0}{
            \node[circle, draw=\origColor, fill=\customColor,
                  minimum size=0.5mm, scale=0.3] at (\plotStartX + \i * \plotWidth + \i * \plotOff + \legendXMarkerStart + \l * 0.3 * \plotWidth + \legendXOff, \plotStartY+ \plotHeight+ \legendYOff+ \legendYOffText){};
        }{
            \node[star, star points=5, star point ratio=0.4, rotate=-35, draw=\origColor, fill=\customColor,
            minimum size=0.5mm, scale=0.5] at (\plotStartX + \i * \plotWidth + \i * \plotOff + \legendXMarkerStart + \l * 0.3 * \plotWidth + \legendXOff, \plotStartY+ \plotHeight+ \legendYOff+ \legendYOffText){};
        }
        \node[anchor=mid west] at (\plotStartX + \i * \plotWidth + \i * \plotOff + \legendXStart + \l * 0.3 * \plotWidth + \legendXOff, \plotStartY+ \plotHeight+ \legendYOff+ \legendYOffText) {%
            \footnotesize \pipelineVersion
            };
    }

            
} 
    
\end{tikzpicture}

%% file: images/violin/violin.tikz
\begin{tikzpicture}

\newcommand\plotStartX{0mm}
\newcommand\plotStartY{0mm}
\newcommand\plotHeight{2.5cm}
\newcommand\plotWidth{3.5cm}
\newcommand\plotOff{3mm}
\newcommand\majorTick{0.75mm}
\newcommand\minorTick{0.5mm}
\newcommand\tickXDist{6.75mm}
\newcommand\tickXOffY{-6mm}
\newcommand\tickXOffX{-4mm}
\newcommand\tickRotation{45}
\newcommand\legendXOff{1mm}
\newcommand\legendXStart{3mm}
\newcommand\legendHeight{3.5mm}
\newcommand\legendYOff{1.25mm}

\foreach \i / \j in {0/1, 1/0}{
    \ifthenelse{\j = 0}{
        \def\metric{pesq}
        \def\ylabel{$\Delta$\acs{pesq}\,$\rightarrow$}
        \def\description{The \textit{speech quality} of \\[0pt] our proposed method is \\[-2pt] preferable to the baseline.}
    }{
        \def\metric{sisdr}
        \def\ylabel{$\Delta$\acs{sisdr}\,[dB]\,$\rightarrow$}
        \def\description{The \textit{suppression of the} \\[0pt] \textit{interfering speaker} of our \\[-2pt] proposed method is superior.}
    }
    \node[anchor=center] at (\plotStartX + 0.5 * \plotWidth, \plotStartY + 0.5 * \plotHeight + \i * \plotHeight + \i * \plotOff) {%
        \pgfimage[height=\plotHeight, width=\plotWidth]{images/violin/delta_\metric.pdf} 
    };

    \ifthenelse{\j = 0}{
        \pgfmathsetlengthmacro{\tickYDist}{\plotHeight / 8}
        \foreach \y in {0, ..., 8}{
            \pgfmathtruncatemacro\tmp{int(\y/2) * 2}
            \pgfmathsetmacro{\yticklabel}{\y * 0.25 - 0.25}
            \ifnum\tmp=\y
                \pgfmathsetmacro{\tickLen}{\minorTick}
            \else
                \pgfmathsetmacro{\tickLen}{\majorTick}
                \node[anchor=east] at (\plotStartX, \plotStartY + \i * \plotHeight + \i * \plotOff + \y * \tickYDist) {\footnotesize \yticklabel};
            \fi
            \draw[line width=\tickWidth] (\plotStartX, \plotStartY + \i * \plotHeight + \i * \plotOff + \y * \tickYDist) --  (\plotStartX - \tickLen, \plotStartY + \i * \plotHeight + \i * \plotOff + \y * \tickYDist);
        }
        \node[anchor=center, rotate=90, yshift=7mm] at (\plotStartX, \plotStartY + \i * \plotHeight + \i * \plotOff + 0.5 * \plotHeight) {\footnotesize \ylabel};
    }{
       \pgfmathsetlengthmacro{\tickYDist}{\plotHeight / 8}
        \foreach \y in {0, ..., 8}{
            \pgfmathtruncatemacro\tmp{int(\y/2) * 2}
            \pgfmathsetmacro{\yticklabel}{int(\y * 2 + 2)}
            \ifnum\tmp=\y
                \pgfmathsetmacro{\tickLen}{\majorTick}
                \node[anchor=east] at (\plotStartX, \plotStartY + \i * \plotHeight + \i * \plotOff + \y * \tickYDist) {\footnotesize \yticklabel};
            \else
                \pgfmathsetmacro{\tickLen}{\minorTick}
            \fi
            \draw[line width=\tickWidth] (\plotStartX, \plotStartY + \i * \plotHeight + \i * \plotOff + \y * \tickYDist) --  (\plotStartX - \tickLen, \plotStartY + \i * \plotHeight + \i * \plotOff + \y * \tickYDist);
        }
        \node[anchor=center, rotate=90, yshift=7mm] at (\plotStartX, \plotStartY + \i * \plotHeight + \i * \plotOff + 0.5 * \plotHeight) {\footnotesize \ylabel};
    }
    
    \pgfmathsetlengthmacro{\tickXDistOff}{
        \plotWidth / 4
    }
    \foreach \x / \xlabel in {
        0 /  {\acs{pf}-\acs{rw}},
        1 /  {\acs{pf}-\acs{cv}}
    }{
        \pgfmathsetmacro{\tickLen}{\majorTick}
        \draw[line width=\tickWidth] (\plotStartX + 2*\x * \tickXDistOff + \tickXDistOff, \plotStartY + \i * \plotHeight + \i * \plotOff) --  (\plotStartX + 2*\x * \tickXDistOff + \tickXDistOff,\plotStartY + \i * \plotHeight + \i * \plotOff - \tickLen );

        \ifthenelse{\i = 0}{
            \node[anchor=north] at (\plotStartX + 2*\x * \tickXDistOff + \tickXDistOff, \plotStartY) {\footnotesize \xlabel};
        }{}
    }

    \ifthenelse{\i = 1}{
        \draw[draw, rounded corners=\rectEdge, line width=\tickWidth]
            (\plotStartX, \plotStartY+ \plotHeight+  \i * \plotHeight + \i * \plotOff + \legendYOff) rectangle (\plotStartX + \plotWidth, \plotStartY+ \plotHeight+  \i * \plotHeight + \i * \plotOff + \legendYOff + \legendHeight);
        \foreach \l / \legendName in {
            0/{concat.}, 1/{\acs{ar}\,(ours)}
        }{
            \node[anchor=mid west] at (\plotStartX + \legendXStart + \l * 0.5 * \plotWidth + \legendXOff, \plotStartY+ \plotHeight+  \i * \plotHeight + \i * \plotOff + \plotOff) {%
                \footnotesize \legendName
                };

            \ifthenelse{\l = 0}{
                \def\origColor{orig_green}
                \def\customColor{custom_green}
            }{
                \def\origColor{orig_purple}
                \def\customColor{custom_purple}
            }
            \node[rectangle, fill=\customColor,
                  minimum size=0.5mm, scale=0.8] at (\plotStartX + \legendXStart + \l * 0.5 * \plotWidth, \plotStartY+ \plotHeight+  \i * \plotHeight + \i * \plotOff + \plotOff){};
                
        }
    }{}
    
} 
    
\end{tikzpicture}

%% file: images/preference_test/preference_test.tikz
\begin{tikzpicture}

\newcommand\plotStartX{0mm}
\newcommand\plotStartY{0mm}
\newcommand\plotHeight{2.5cm}
\newcommand\plotWidth{3.5cm}
\newcommand\plotOff{3mm}
\newcommand\majorTick{0.75mm}
\newcommand\minorTick{0.5mm}
\newcommand\tickXDist{6.75mm}
\newcommand\tickXOffY{-6mm}
\newcommand\tickXOffX{-4mm}
\newcommand\tickRotation{45}
\newcommand\descriptionOff{2mm}
\newcommand\legendXOff{1mm}
\newcommand\legendXStart{3mm}
\newcommand\legendHeight{3.5mm}
\newcommand\legendYOff{1.25mm}

\foreach \i / \j in {0/1, 1/0}{
    \pgfmathsetmacro{\plotIdx}{int(\j)}

    \node[anchor=center] at (\plotStartX + 0.5 * \plotWidth, \plotStartY + 0.5 * \plotHeight + \i * \plotHeight + \i * \plotOff) {%
        \pgfimage[height=\plotHeight, width=\plotWidth]{images/preference_test/preference_test_\plotIdx.pdf} 
    };

    \ifthenelse{\j = 0}{
        \def\description{The \textit{speech quality} of \\[0pt] our proposed method is \\[-2pt] preferable to the baseline.}
    }{
        \def\description{The \textit{suppression of the} \\[0pt] \textit{interfering speaker} of our \\[-2pt] proposed method is better.}
    }
    \newcommand\backH{1cm}
    \pgfmathsetlengthmacro{\backW}{0.925 * \plotWidth}
    \draw[fill=white, fill opacity=0.7, draw=none] 
  (\plotStartX+ \descriptionOff - 0.5mm, \plotStartY + 1 * \plotHeight + \i * \plotHeight + \i * \plotOff - 0.75mm) rectangle ++(\backW, -\backH);
    \node[text width=\plotWidth, align=left, anchor=north] at (\plotStartX + 0.5 * \plotWidth + \descriptionOff, \plotStartY + 1 * \plotHeight + \i * \plotHeight + \i * \plotOff) {%
        \footnotesize \enquote{\description}
    }; 

    \pgfmathsetlengthmacro{\tickYDist}{\plotHeight / 10}
    \foreach \y in {0, ..., 10}{
        \pgfmathtruncatemacro\tmp{int(\y/2) * 2}
        \pgfmathsetmacro{\ylabel}{int(\y * 10)}
        \ifnum\tmp=\y
            \pgfmathsetmacro{\tickLen}{\majorTick}
            \node[anchor=east] at (\plotStartX, \plotStartY + \i * \plotHeight + \i * \plotOff + \y * \tickYDist) {\footnotesize \ylabel};
        \else
            \pgfmathsetmacro{\tickLen}{\minorTick}
        \fi
        \draw[line width=\tickWidth] (\plotStartX, \plotStartY + \i * \plotHeight + \i * \plotOff + \y * \tickYDist) --  (\plotStartX - \tickLen, \plotStartY + \i * \plotHeight + \i * \plotOff + \y * \tickYDist);
    }
    \node[anchor=center, rotate=90, yshift=7mm] at (\plotStartX, \plotStartY + \i * \plotHeight + \i * \plotOff + 0.5 * \plotHeight) {\footnotesize participants\,[\%]};
    
    \pgfmathsetlengthmacro{\tickXDistOff}{
        0.5 * (\plotWidth - 4 * \tickXDist)
    }
    \foreach \x / \xlabel in {
        0 /  \parbox{2cm}{\centering strongly \\[-2pt] disagree},
        1 /  \parbox{2cm}{\centering disagree},
        2 /  \parbox{2cm}{\centering \hspace{1.5mm}neutral},
        3 /  \parbox{2cm}{\centering \hspace{3mm}agree},
        4 /  \parbox{2cm}{\centering strongly \\[-2pt] agree}
    }{
        \pgfmathsetmacro{\tickLen}{\majorTick}
        \draw[line width=\tickWidth] (\plotStartX + \x * \tickXDist + \tickXDistOff, \plotStartY + \i * \plotHeight + \i * \plotOff) --  (\plotStartX + \x * \tickXDist + \tickXDistOff,\plotStartY + \i * \plotHeight + \i * \plotOff - \tickLen );

        \ifthenelse{\i = 0}{
            \node[rotate=\tickRotation, align=center, anchor=center] at (\plotStartX + \x * \tickXDist + \tickXDistOff + \tickXOffX, \plotStartY + \tickXOffY) {\footnotesize \xlabel};
        }{}
    }

    \ifthenelse{\i = 1}{
    \draw[draw, rounded corners=\rectEdge, line width=\tickWidth]
            (\plotStartX, \plotStartY+ \plotHeight+  \i * \plotHeight + \i * \plotOff + \legendYOff) rectangle (\plotStartX + \plotWidth, \plotStartY+ \plotHeight+  \i * \plotHeight + \i * \plotOff + \legendYOff + \legendHeight);
        \foreach \l / \legendName in {
            0/{\acs{pf}-\acs{rw}}, 1/{\acs{pf}-\acs{cv}}
        }{
            \node[anchor=west] at (\plotStartX + \legendXStart + \l * 0.5 * \plotWidth + \legendXOff, \plotStartY+ \plotHeight+  \i * \plotHeight + \i * \plotOff + \plotOff) {%
                \footnotesize \legendName
                };
            \ifthenelse{\l = 0}{
                \node[circle, draw=dark_gray, fill=white,
                  minimum size=0.5mm, scale=0.3] at (\plotStartX + \legendXStart + \l * 0.5 * \plotWidth, \plotStartY+ \plotHeight+  \i * \plotHeight + \i * \plotOff + \plotOff){};
            }{
                \node[star, star points=5, star point ratio=0.4, rotate=-35, draw=dark_gray, fill=white,
                minimum size=0.5mm, scale=0.5] at (\plotStartX + \legendXStart + \l * 0.5 * \plotWidth, \plotStartY+ \plotHeight+  \i * \plotHeight + \i * \plotOff + \plotOff){};
            }
        }
    }{}
} 
    
\end{tikzpicture}